\documentclass[aps,prb,twocolumn,superscriptaddress]{revtex4-1}

\usepackage{physics}
\usepackage{bbold}
\usepackage{lgrind}        
\usepackage{natbib}    
\usepackage{color}         
\usepackage[pdftex]{graphicx}      
\usepackage{longtable}     
\usepackage{epsf}          
\usepackage{bm}            
\usepackage{amsmath}
\usepackage{thumbpdf}
\usepackage{esint}
\usepackage{subfigure}
\usepackage[colorlinks=true]{hyperref} 
\begin{document}


\title{Prediction of $\mathrm{TiRhAs}$ as a Dirac Nodal Line Semimetal via First-Principles Calculations}


\author{Sophie F. Weber}
\affiliation{Department of Physics, University of California, Berkeley, CA 94720, USA}
\affiliation{Molecular Foundry, Lawrence Berkeley National Laboratory, Berkeley, CA 94720, USA}
\author{Ru Chen}
\affiliation{Department of Physics, University of California, Berkeley, CA 94720, USA}
\affiliation{Molecular Foundry, Lawrence Berkeley National Laboratory, Berkeley, CA 94720, USA}
\author{Qimin Yan}
\affiliation{Department of Physics, Temple University, Philadelphia, PA 19122, USA}
\author{Jeffrey B. Neaton}
\affiliation{Department of Physics, University of California, Berkeley, CA 94720, USA}
\affiliation{Molecular Foundry, Lawrence Berkeley National Laboratory, Berkeley, CA 94720, USA}


\date{\today}

\begin{abstract}
Using first-principles calculations we predict that $\mathrm{TiRhAs}$, a previously synthesized compound, is a Dirac nodal line (DNL) semimetal. The DNL in this compound is found to be protected both by the combination of inversion and time-reversal symmetry, and by a reflection symmetry, in the absence of spin-orbit coupling (SOC). Our calculations show that band velocities associated with the nodal line have a high degree of directional anisotropy, with in-plane velocities $v_\perp$ perpendicular to the nodal line between $1.2-2.8\times10^5$ m/s. The crossings along the DNL are further found to exhibit a prominent and position-dependent tilt along directions perpendicular to the nodal line. We calculate $\mathbb{Z}_2$ indices based on parity eigenvalues at time-reversal invariant momenta and show that $\mathrm{TiRhAs}$ is topological. A tight-binding model fit from our first-principles calculations demonstrates the existence of two-dimensional drumhead surface states on the surface Brillouin zone. Based on the small gapping of the DNL upon inclusion of SOC and the clean Fermi surface free from trivial bands, $\mathrm{TiRhAs}$ is a promising candidate for further studies of the properties of topological semimetals. 
\end{abstract}


\maketitle

\section{Introduction}
\indent A recent development in the field of condensed matter physics is the discovery of topological semimetals (TSMs)\cite{Weng2016,Murakami2007}. These materials have robust, symmetry-protected crossings in reciprocal space, and can be characterized by topological invariants, analogous to the topological insulators (TIs). Three types of TSMs which have been studied in detail both theoretically and experimentally are\cite{Weng2016}Weyl semimetals, hosting pairs of massless twofold degenerate nodal points with opposite chirality in the three-dimensional (3D) Brillouin zone (BZ)\cite{Xu2015,LV2015,Wan2011,Weng2015a}; Dirac semimetals, with fourfold degenerate nodal points consisting of overlapping Weyl points\cite{Liu2014a,Liu2014,Wang2013}; and Dirac nodal line semimetals (DNLs), in which the valence and conduction bands touch in a closed loop in momentum space\cite{Hu2016,Bian2016a,Neupane2016,Yu2015}. All three categories are expected to display unusual and intriguing properties, such as ultrahigh mobility, giant magnetoresistance, chiral anomalies, and surface states\cite{Liang2014,Parameswaran2014}.\\
\indent DNLs are unique from other types of TSMs by virtue of having a one-dimensional Fermi surface, in contrast to the zero-dimensional Fermi surfaces of Weyl and Dirac semimetals. This implies that the density of states (DOS) of low-energy bulk excitations is quadratic in $\abs{E-E_f}$, where $E_f$ is the Fermi energy, rather than linear\cite{Bian2016b}. The larger DOS means that interaction-induced instabilities which are predicted for Weyl semimetals can be even more pronounced in DNLs\cite{Bian2016a}. The one-dimensional nature of the Fermi surface also suggests that such compounds may exhibit effects from long-range Coulumb interactions due to reduced screening\cite{Huh2016}. Finally, the topological surface states of DNLs, which take the form of a two-dimensional "drumhead" terminating on the projection of the nodal line onto the surface BZ, have been suggested to provide a platform for exotic physics arising from electronic correlations\cite{Chan2016}.\\ 
\indent In spite of their numerous desirable properties, less than ten DNL compounds have been identified or verified experimentally thus far\cite{Burkov2011,Hu2016}. This might seem surprising given the fact that there are many different crystalline symmetries that that can stabilize a DNL. But a challenge to experimental realization is that the majority of these protecting mechanisms are only robust when spin-orbit coupling (SOC) is ignored\cite{Fang2016}; the nodal line degeneracies are often lifted to a significant degree by SOC unless an additional nonsymmorphic symmetry, such as a screw axis, is present\cite{Fang2015,Chen2017}. Another experimental difficulty for DNL compounds, including those synthesized thus far is that there are often trivial bulk bands near the Fermi level coexisting and interfering with the nontrivial nodal line, making a definitive experimental study of the topological properties challenging\cite{Neupane2016}.\\
\indent Here, we use first-principles calculations to predict that $\mathrm{TiRhAs}$, which has been synthesized in the past\cite{Roy-Montreuil1984} but whose electronic properties have thus far remained unexamined, is a DNL semimetal with a nodal line around the Fermi energy which lies in the $k_x=0$ plane, pinned to the plane by a mirror symmetry. Our study of $\mathrm{TiRhAs}$ is motivated by several factors. First, its nonsymmorphic space group possesses several of the symmetry elements known to protect nodal lines. Second, the effect of SOC, given the elements involved, is likely small, and the lifting of degeneracy is expected to be nearly negligible. Lastly, because $\mathrm{TiRhAs}$ has an even number of electrons per unit cell, the Kohn-Luttinger theorem suggests any DNL might be fixed near the Fermi energy\cite{Luttinger1960,Wan2011}, which is desirable for further experimental study and for applications.\\
\section{Results}
\subsection{Crystal structure and methodology}
\begin{figure}
\includegraphics[width=\columnwidth]{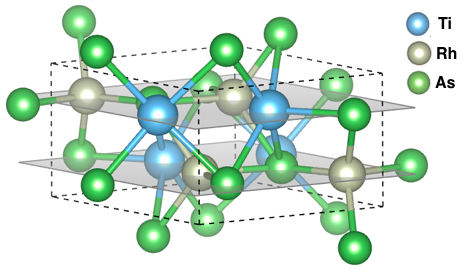}
\caption{\label{fig:struct}Orthorhombic crystal structure of $\mathrm{TiRhAs}$, with space group \emph{Pmnb}. The primitive cell consists of two mirror planes perpendicular to $(100)$, each with two Ti atoms (blue), two As atoms (green), and two Rh atoms (gray).}
\end{figure}

\indent Prior experimental work has shown that $\mathrm{TiRhAs}$ crystallizes in an orthorhombic lattice with the nonsymmorphic centrosymmetric space group \emph{Pmnb} [62]\cite{Roy-Montreuil1984}. The primitive cell is shown in Figure \ref{fig:struct}. It is composed of two layers with six atoms each in the $(\frac{1}{4},y,z)$ and $(\frac{3}{4},y,z)$ planes. Each Ti atom is five-fold coordinated by As in the shape of "distorted" edge-sharing square pyramids; the Rh atoms are tetragonally-coordinated by As.\par
For our first-principles density functional theory (DFT) calculations on $\mathrm{TiRhAs}$, we use the Vienna \emph{ab initio} simulation package (VASP)\cite{Kresse1996}with generalized gradient approximation (GGA) using the Perdew-Burke-Ernzerhof (PBE) functional\cite{Perdew1996} and projector augmented-wave method (PAW)\cite{Blochl1994}. The PAW-PBE pseudopotentials of Ti, Rh and As treat $3d^24s^2$, $4d^85s^1$ and $4s^24p^3$ electrons as valence states. We employ an energy cutoff of 300 eV for our plane wave basis set and a Monkhorst-Pack $\mathbf{k}$-point mesh of $8\times6\times6$. Brillouin zone integrations are performed with a Gaussian broadening of 0.05 eV during all calculations\cite{Elsasser1994}. These parameters lead to total energies converged to within a few meV. We fully relax the lattice parameters starting from the experimental values. Our GGA lattice constants agree with the experimental results to within $1\%$ (see Table \ref{tab: params}). We use the optimized lattice parameters for all band structure calculations. For calculations with SOC, we include SOC self-consistently\cite{Theurich2001}.  

 \begin{table}
 \caption{\label{tab: params}Comparison between experimental lattice parameters and Wyckoff positions, and values obtained after full optimization with DFT-PBE.}
 \begin{ruledtabular}
 \begin{tabular}{| c | c | c |}
 \hline
 & Experiment\cite{Roy-Montreuil1984}& DFT-PBE\\ \hline
$a$ (\AA) & $3.816$ & $3.841$ \\
$b$ (\AA)& $6.334$ & $6.366$\\
$c$ (\AA) & $7.388$ & $7.434$  \\ \hline
 Rh (4c) y & $0.855$ & $0.857$ \\
 Rh (4c) z & $0.064$ & $0.063$\\ \hline
 Ti (4c) y& $0.972$ & $0.968$ \\
 Ti (4c) z& $0.684$ & $0.682$\\ \hline
 As (4c) y& $0.243$ & $0.252$ \\
 As (4c) z& $0.122$ & $0.122$ \\
 \hline
\end{tabular}
\end{ruledtabular}
\end{table}
\subsection{Band structure and symmetries}
\begin{figure}
\subfigure[]{\label{fig:orbital}}\includegraphics{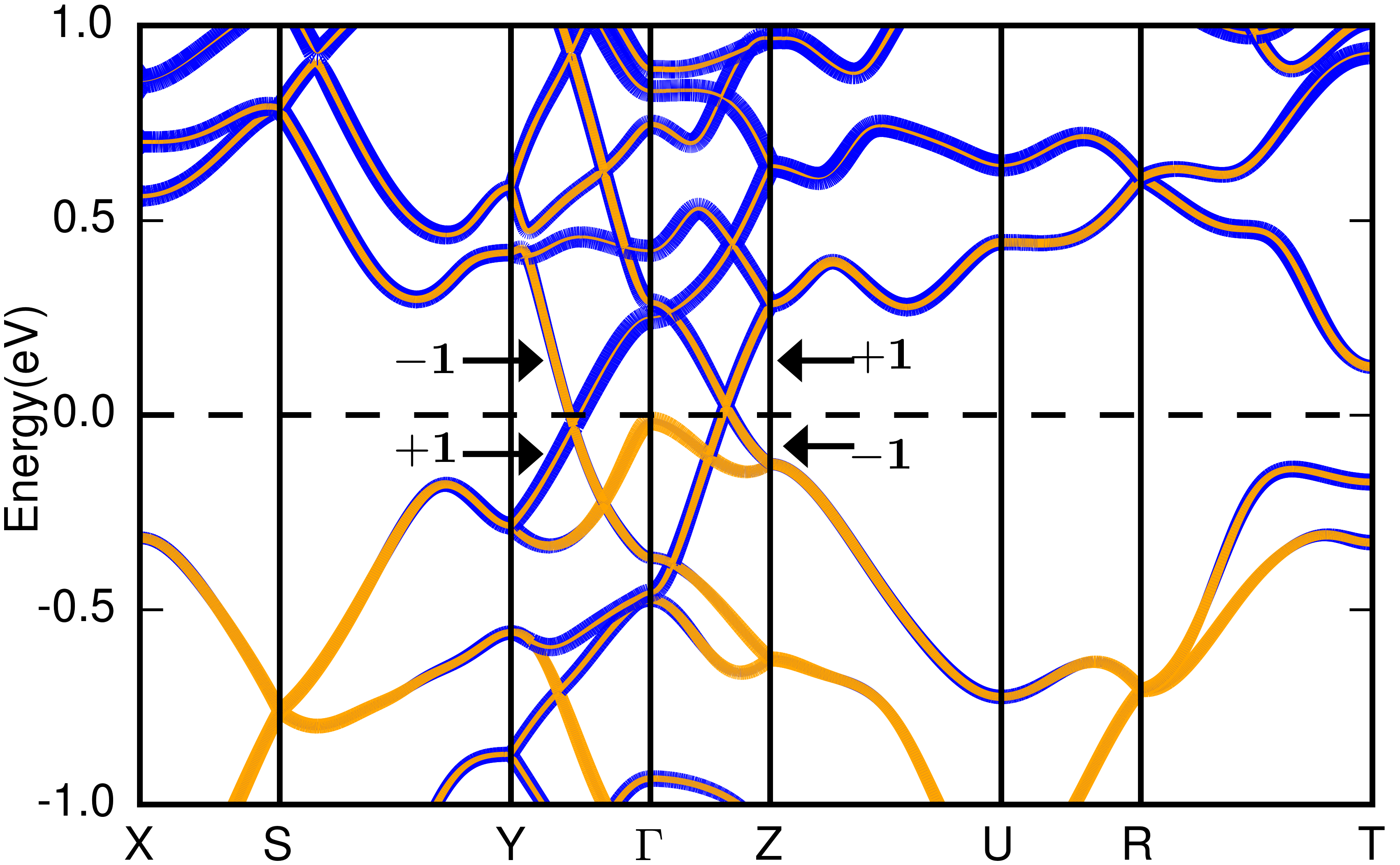}
\subfigure[]{\label{fig:SOC}}\includegraphics{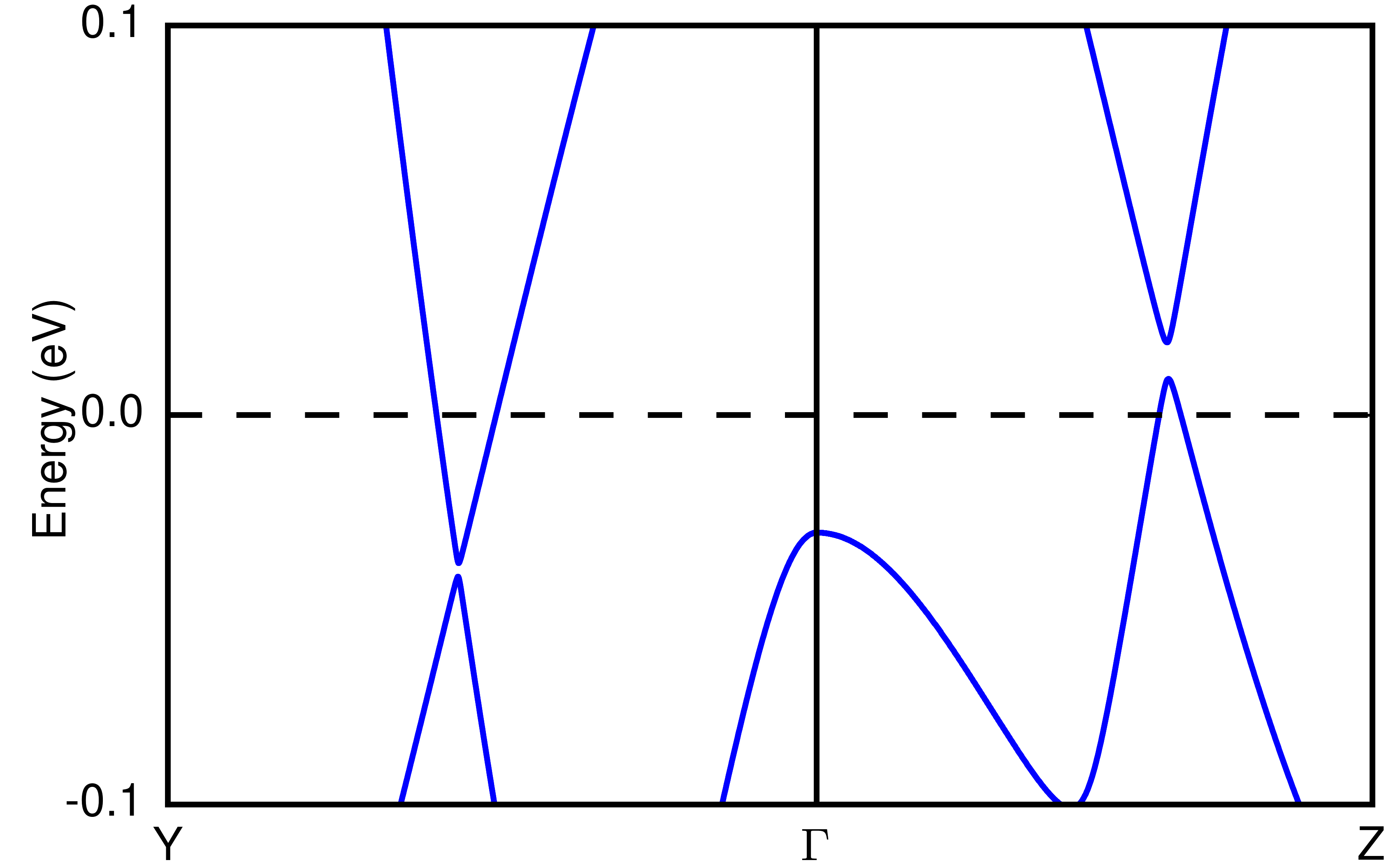}
\caption{(a)DFT-PBE band structure without SOC. The projection of the bands onto Ti $d$ orbitals is shown in blue, and the projection onto Rh $d$ orbitals is shown in orange. The widths of the lines are proportional to the values. The mirror eigenvalues $\pm{1}$ of the crossing bands are also indicated. (b)DFT-PBE band structure with SOC included, indicating that SOC opens a small gap due to hybridization of bands with like mirror eigenvalues.}
\label{fig:bandstructs}
\end{figure}

\indent The GGA band structure of $\mathrm{TiRhAs}$ without SOC is plotted in Figure \ref{fig:orbital}. The band crossings along the high symmetry lines $Y-\Gamma$ and $\Gamma-Z$ indicate the presence of a nodal line in the $k_x=0$ plane encircling $\Gamma$. From an analysis of the site and angular momentum-projected band character, we find that the bands near the crossings are a mixture of Rh 4\emph{d} and Ti 3\emph{d} states. Because GGA is known to overestimate band inversion\cite{Vidal2011}, we also compute bulk band structures using the hybrid density functional HSE06\cite{Heyd2003}. The HSE06 result reproduces the DNL and yields an even cleaner Fermi surface than GGA, as the lone trivial band at $\Gamma$ is pushed down relative to $E_f$ to lower energies (see Supplementary Material\cite{suppmat}).\\
\indent The DNL in $\mathrm{TiRhAs}$ is protected by two different symmetries: (a) the combination of inversion and time reversal symmetries $\mathcal{P}$ and $\mathcal{T}$ in the absence of SOC, and (b) a mirror plane at $x=\frac{a}{4}$. The protection of DNLs by $\mathcal{P}$ and $\mathcal{T}$, provided that SOC is ignored, has been discussed extensively in the literature\cite{Chan2016,Kim2015,Fang2015,Yu2015,Weng2015,Weng2016,Huang2016}. Here we briefly motivate why the generic solution for a band crossing in such a system is a closed nodal line (rather than discrete crossings) using a codimension argument\cite{Huang2016}. The Bloch Hamiltonian $\mathcal{H}(\mathbf{k})$ for a spinless system near a generic band crossing may always be written as a linear combination of the identity and the three Pauli matrices, with $\mathbf{k}$-dependent coefficients. The combination of $\mathcal{P}$ and $\mathcal{T}$ allow us to choose a gauge for the cell-periodic part $u_{n\mathbf{k}}(r)$ of the Bloch eigenfunctions in which $u_{n\mathbf{k}}^*(r)=u_{n\mathbf{k}}(-r)$. From this fact it trivially follows that $\mathcal{H}(\mathbf{k})$ is real-valued. Consequently we can always express $\mathcal{H}(\mathbf{k})$ in terms of only two of the three Pauli matrices, giving our band crossing a codimension of two. Since this is one less than the number of independent variables ($k_x$,$k_y$,$k_z$), the generic solution $E(\mathbf{k})$ is a line node, which will always be stable in the presence of $\mathcal{P}$ and $\mathcal{T}$. \\
\indent We now discuss the consequences of the mirror symmetry $\mathcal{R}_x$ in the absence of SOC. The explicit form of the operator in real space is
\begin{equation}
\mathcal{R}_x: (x,y,z)\rightarrow(-x+\frac{a}{2},y,z).
\label{eq:realspacemirror}
\end{equation}
It is clear from equation \ref{eq:realspacemirror} that $\mathcal{R}_x^2=+1$. Thus the eigenvalues of $\mathcal{R}_x$ in the absence of SOC are $\pm{1}$. Moreover, the action of $\mathcal{R}_x$ in reciprocal space is
\begin{equation}
\mathcal{R}_x: (k_x,k_y,k_z)\rightarrow(-k_x,k_y,k_z).
\label{eq:recspacemirror}
\end{equation}
 Therefore, all Bloch functions $\psi_{n\mathbf{k}}(\mathbf{r})=e^{i\mathbf{k}\cdot{\mathbf{r}}}u_{n\mathbf{k}}(\mathbf{r})$ in the $k_x=0$ plane are invariant under $\mathcal{R}_x$, meaning that the bands in this plane may be labeled by the mirror eigenvalues $\pm{1}$. Bands with the same mirror eigenvalue can hybridize, leading to a band gap. However, bands with opposite eigenvalues are symmetry-forbidden from mixing and thus their crossing is protected. We check the $\mathcal{R}_x$ eigenvalues of the valence and conduction bands along $Y-\Gamma-Z$ using wavefunctions obtained with the all-electron WIEN2k code\cite{Blaha2001} and confirm that the crossing bands have opposite eigenvalues, as shown in Figure \ref{fig:orbital} (details of our WIEN2k calculations appear in the Supplementary Material). It should be noted that even if the mirror symmetry is broken in $\mathrm{TiRhAs}$, the DNL will still be protected as long as $\mathcal{P}$ and $\mathcal{T}$ symmetries persist; it will merely be unpinned from the $k_x=0$ plane.\\
\indent The self-consistent inclusion of SOC opens a small gap (Figure \ref{fig:SOC}). The DFT-PBE-SOC gap varies depending on position along the DNL from less than $1$ meV to a maximum of $40$ meV. SOC gaps the nodal line by coupling spin and spatial degrees of freedom. Thus, $\mathcal{R}_x$ not only maps $x$ to $\frac{1}{2}-x$, but also maps $s_{y,z}$ to $-s_{y,z}$, i.e, the effect of $\mathcal{R}_x$ on spin space is to perform a $\pi$ rotation about the $\hat{x}$ axis (The difference in the effect of $\mathcal{R}_x$ on real space and spin space is due to the fact that spin is a pseudovector). Now, with SOC included, squaring $\mathcal{R}_x$ amounts to a $2\pi$ rotation in spin space which gives a minus sign for a spin-$\frac{1}{2}$ system, meaning that the eigenvalues of $\mathcal{R}_x$ become $\pm{i}$. Thus, each band with eigenvalue $\pm{1}$ in the non-SOC system becomes doubly degenerate with mirror eigenvalues $\pm{i}$ in the SOC system. Conduction and valence bands with the same eigenvalues can now hybridize, leading to an anticrossing (Figure \ref{fig:SOC}). We wish to emphasize that while absence of SOC is necessary to keep conduction and valence bands completely degenerate along the DNL, $\mathrm{TiRhAs}$ maintains its nontrivial $\mathbb{Z}_2$ indices even with SOC\cite{Kim2015}.\\
\indent Finally, we check the DFT-PBE band structure upon several isovalent substitutions for $\mathrm{TiRhAs}$, specifically $\mathrm{TiCoAs}$, $\mathrm{TiRhP}$ and $\mathrm{ZrRhAs}$. We start with the optimized lattice parameters of $\mathrm{TiRhAs}$ and relax these substituted structures within the \emph{Pmnb} space group. The resulting lattice parameters in all three cases deviate from the starting values by {$0.4$\AA}  at most. The band structures are qualitatively identical to $\mathrm{TiRhAs}$; in particular, they all have a DNL in the $k_x=0$ plane. This implies that as long as the valence electron count is preserved, partial or full isovalent substitution may be attempted in order to reduce or enhance the effect of SOC. \\
\subsection{$\mathbf{k}\cdot{\mathbf{p}}$ analysis of band velocities}
\indent A detailed analysis of  band velocities at various points along the DNL (where the band velocity $v_{n\mathbf{k}}$ is given by $\frac{1}{\hbar} \frac{\partial{E_{n\mathbf{k}}}}{\partial{\mathbf{k}}}$ at the crossing point of interest), is crucial for understanding transport experiments. We employ a $\mathbf{k}\cdot{\mathbf{p}}$ analysis to ascertain the symmetry constraints in $\mathrm{TiRhAs}$ that determine the $\mathbf{k}$-dependent band velocities along the nodal line. The generators of the space group \emph{Pmnb} are two mutually perpendicular two-fold screw axes and inversion $\mathcal{P}$. Since $\mathrm{TiAsRh}$ is nonmagnetic, $\mathcal{T}$ is also a symmetry as discussed above. The Bloch Dirac Hamiltonian (without SOC) may be expanded around any point $\mathbf{k}$ on the DNL as
\begin{figure*}
\subfigure[]{\label{fig:DNL}}\includegraphics[width=.67\columnwidth]{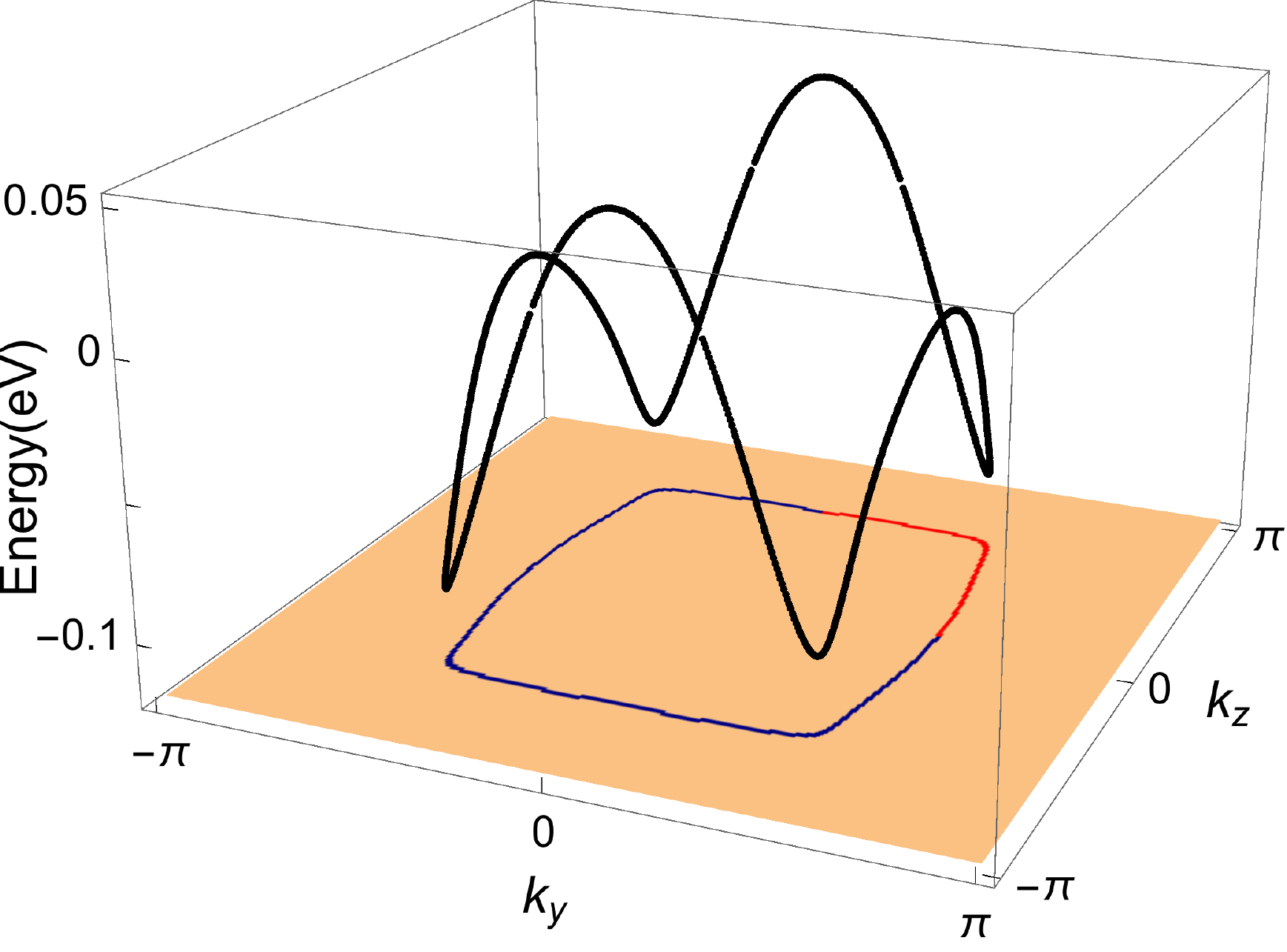}.
\subfigure[]{\label{fig:speed}}\includegraphics{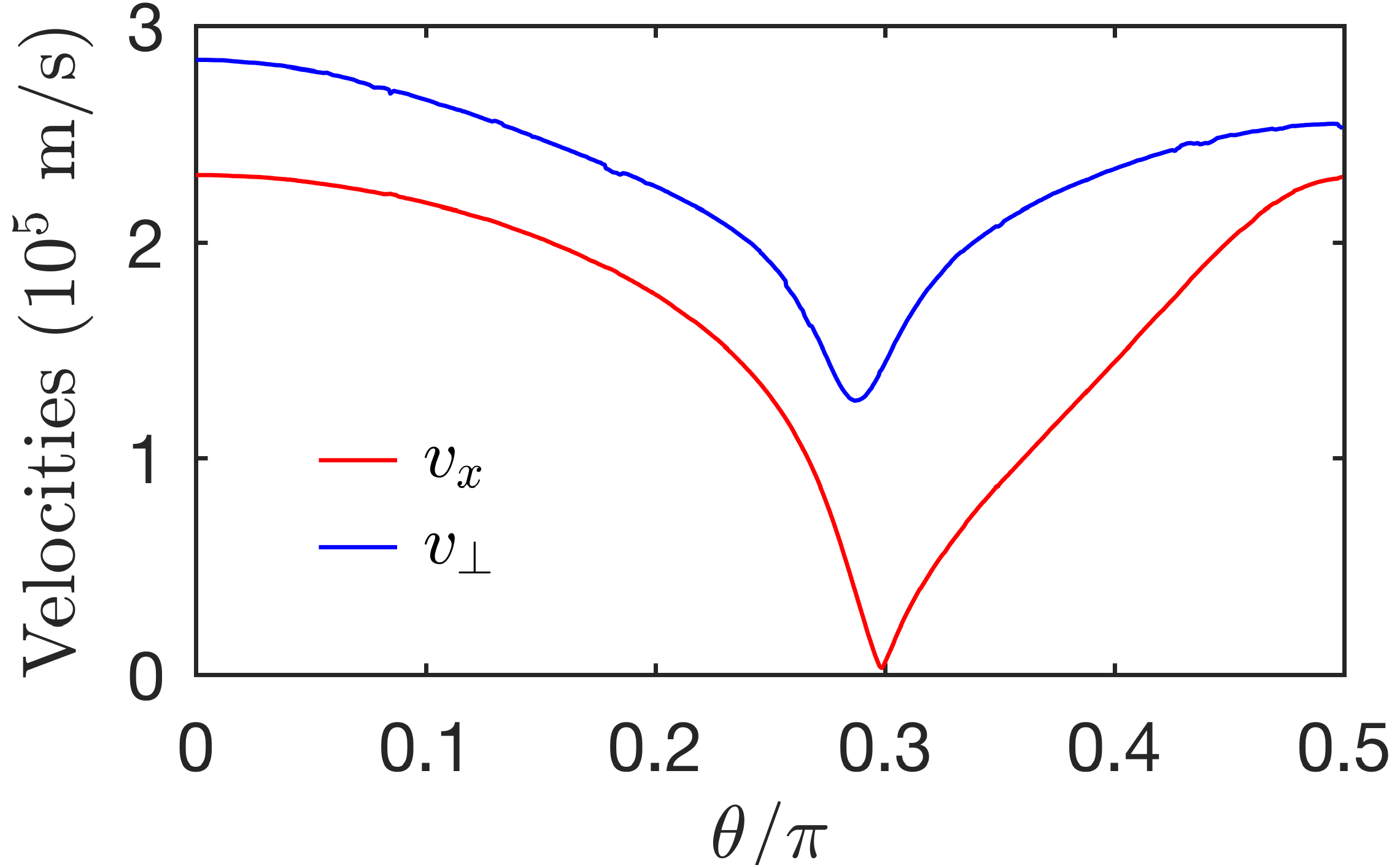}
\subfigure[]{\label{fig:tilt}}\includegraphics{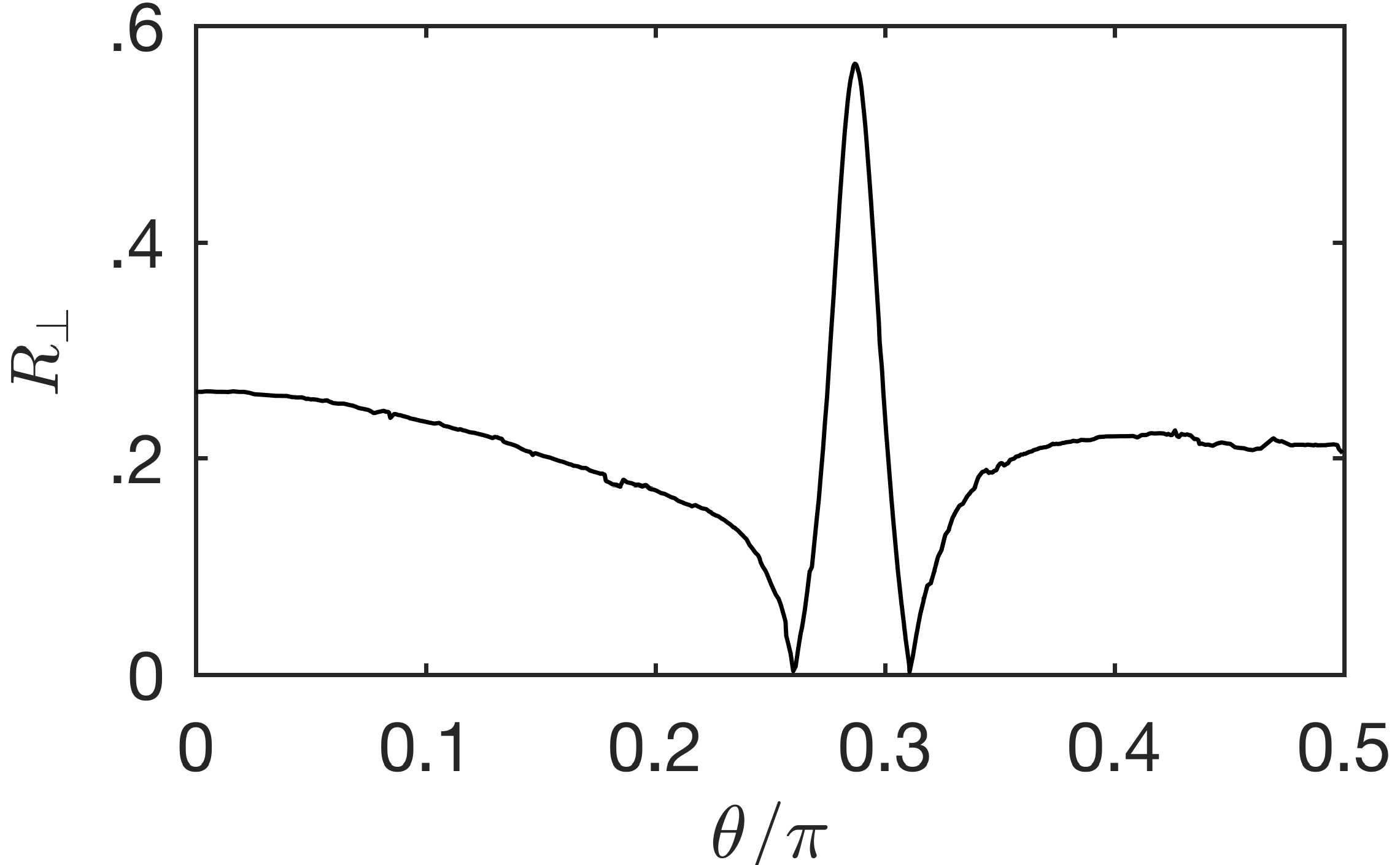}
\caption{(a) DFT-PBE calculations of the DNL in $\mathrm{TiRhAs}$ in the $k_x=0$ plane, with the irreducible quadrant highlighted in red. (b) $v_\perp$ and $v_x$ as a function of $\theta$. (c) Ratio of tilting magnitude to isotropic velocity, $R_\perp$, along the $v_\perp$ direction as a function of $\theta$.}
\label{fig:kdotp}
\end{figure*}
\begin{equation}
\mathcal{H}(\mathbf{k}+\delta{\mathbf{k}})=E(\mathbf{k})+h_\mu^i(\mathbf{k})\sigma^{\mu}\delta{k_i}+\mathcal{O}(\delta\mathbf{k}^2),
\label{eq:kdotp}
\end{equation}
\noindent where $\delta\mathbf{k}=(\delta k_x,\delta k_y, \delta k_z)$ is the deviation from a point $\mathbf{k}$ in the Brillouin zone, $\sigma^{\mu}$ are the Pauli matrices with $\mu \in{0,1,2,3}$ and $i\in{x,y,z}$ and $h_\mu^i(\mathbf{k})$ are real, $\mathbf{k}$-dependent coefficients.\\
\indent We now restrict our discussion to the $k_x=0$ plane in which the DNL lies. For a generic point on this plane the only remaining space group symmetry is the mirror symmetry $\mathcal{R}_x$. The product of inversion and time reversal, $\mathcal{P}\mathcal{T}$, is also a symmetry. If we choose the two crossing bands with $+1$ and $-1$ $\mathcal{R}_x$ eigenvalues as pseudospin up and down, respectively, the symmetries for the nodal line may be expressed as 
\begin{equation}
\mathcal{R}_x=\sigma^3; \mathcal{P}\mathcal{T}=\sigma^0K,
\label{eq:symreps}
\end{equation}
where $K$ denotes complex conjugation. These symmetries place constraints on the allowed $h_\mu^i(\mathbf{k})$;
\begin{equation}
\mathcal{R}_x\mathcal{H}(k_x,k_y,k_z)\mathcal{R}^{-1}_x=\mathcal{H}(-k_x,k_y,k_z)
\label{eq:Rrestrict}
\end{equation}
and
\begin{equation}
(\mathcal{P}\mathcal{T})\mathcal{H}(k_x,k_y,k_z)(\mathcal{P}\mathcal{T})^{-1}=\mathcal{H}(k_x,k_y,k_z).
\label{eq:PTrestrict}
\end{equation}
It follows from Equations \ref{eq:Rrestrict} and \ref{eq:PTrestrict} that the only nonzero $h_\mu^i(\mathbf{k})$ values are $h_0^{y,z}(\mathbf{k})$, $h_3^{y,z}(\mathbf{k})$, and $h_1^{x}(\mathbf{k})$. The band dispersion at each point $\mathbf{k}$ on the nodal line can then be expressed as
\begin{multline}
\delta E_{\mathbf{k}+\delta \mathbf{k}}\approx h_0^y(\mathbf{k})\delta k_y+h_0^z(\mathbf{k})\delta k_z\\
\pm\sqrt{(h_3^y(\mathbf{k})\delta k_y+h_3^z(\mathbf{k})\delta k_z)^2+(h_1^x(\mathbf{k})\delta k_x)^2},
\label{eq:dispersion}
\end{multline}
where $\delta E_{\mathbf{k}+\delta \mathbf{k}}=E_{\mathbf{k}+\delta \mathbf{k}}-E_{\mathbf{k}}$. \par
\indent We fit Equation \ref{eq:dispersion} to our DFT calculations at each point on the nodal line in the irreducible quadrant of the BZ. The coefficients outside the square root, $h_0^y(\mathbf{k})$ and $h_0^z(\mathbf{k})$, are symmetry-allowed "tilting" terms which characterize the tilting of the Dirac cone along $k_y$ and $k_z$ respectively\cite{Soluyanov2015,Chan2016a,Trescher2015}. The terms inside the square root can be written as $\sum_{i,j=1}^3{\mathcal{A}_{ij}\delta k_i \delta k_j}$ where $\mathcal{A}_{ij}$ is a real symmetric matrix. The square root of the eigenvalues of $\mathcal{A}$ correspond to the principle components of $v_{n\mathbf{k}}$ when the tilt terms are neglected, i.e the splitting of the Dirac cone; they are $(0,\sqrt{(h_3^y(\mathbf{k}))^2+(h_3^z(\mathbf{k}))^2},h_1^x(\mathbf{k}))\propto (v_{\parallel}(\mathbf{k}),v_{\perp}(\mathbf{k}),v_x(\mathbf{k}))$, where $v_{\parallel}$ is tangential to the DNL, $v_{\perp}$ is perpendicular to the DNL in the $k_x=0$ plane, and $v_x$ is along $k_x$. The zero $v_{\parallel}$ corresponds to the "soft" direction where the dispersion scales at least as $\mathcal{O}(\delta \mathbf{k}^2)$. We parametrize points along the DNL by the polar angle $\theta=\tan^{-1}{k_z/k_y}$,  and plot $v_\perp$ and $v_x$ as a function of $\theta$ in Figure \ref{fig:speed}. As shown, our computed DFT-PBE $v_\perp$ is between  $1.2-2.8\times 10^5$ m/s, on the same order of magnitude as reported values for $\mathrm{Na_3Bi}$ and $\mathrm{Cd_3As_2}$\cite{Liu2014,Liang2014}. $v_x$ is computed to be smaller at all $\theta$ and more anisotropic, between $3\times 10^3$ and $2.3\times 10^5$ m/s. \par
\indent From an experimental perspective, while both tilt and relative magnitudes of the velocities given by $\mathcal{A}$ affect directional dependence of conductance in transport experiments, tilt also has an effect on the Fano factor (the ratio of shot noise to current)\cite{Trescher2015}; thus, quantitative characterization is important. The relative degree of tilting at the point $\mathbf{k}$ on the DNL in the direction $(\delta k_x,\delta k_y,\delta k_z)$ is given by the ratio of the magnitude of the tilting to the magnitude of the "isotropic" velocity, which for $\mathrm{TiRhAs}$ is
\begin{equation}
R=\frac{\abs{h_0^y(\mathbf{k})\delta k_y+h_0^z(\mathbf{k})\delta k_z}}{\abs{(h_3^y(\mathbf{k})\delta k_y+h_3^z(\mathbf{k})\delta k_z)^2+(h_1^x(\mathbf{k})\delta k_x)^2}}.
\label{eq:tilt}
\end{equation}
Values greater than 1 indicate a switch in the sign of the the dispersion, analogous to the the type-II Weyl semimetals\cite{Soluyanov2015}.  For concreteness, we choose the direction $(0,h_3^y(\mathbf{k})/h_3^z(\mathbf{k}),1)$ parallel to $v_\perp$ and plot $R_\perp$ as a function of $\theta$ in Figure \ref{fig:tilt}. We see that amount of tilting along $v_\perp$ varies greatly, ranging from nearly $0$ to $.56$ at $\frac{\theta}{\pi}\approx0.3$, at the same point where $v_\perp$ has a prominent dip. 
\subsection{$\mathbb{Z}_2$ invariant}
In order to confirm the topological nature of $\mathrm{TiRhAs}$ and its robustness, we calculate $\mathbb{Z}_2$ invariants analogous to those used to characterize three-dimensional topological insulators (TIs) for systems with inversion symmetry as formulated by Fu and Kane\cite{Fu2007}. The authors showed that in a compound with $\mathcal{P}$ and $\mathcal{T}$ symmetries (and SOC which drives the topological gapping), the topological invariants $(\nu_0;\nu_1\nu_2\nu_3)$ can be computed via the parity eigenvalues $\epsilon_n$ of the occupied Bloch states at the eight time-reversal invariant momenta (TRIM) in the 3D BZ, defined by $\Gamma_i=(n_1\mathbf{b}_1+n_2\mathbf{b}_2+n_3\mathbf{b}_3)/2$, where $n_j=0,1$ and the $\mathbf{b}$ denote the primitive reciprocal lattice vectors. Defining $\epsilon_i=\prod_{n_{occ}}\epsilon_n(\Gamma_i)$, i.e. the product of the parity eigenvalues of all occupied bands at the TRIM point $\Gamma_i$, the "strong" topological index $\nu_0$ is given by
\begin{equation}
(-1)^{\nu_0}=\prod_{i=1}^8\epsilon_i,
\label{eq:strongz2}
\end{equation}
where the product is over the eight TRIM points.  The "weak" indices $\nu_{1,2,3}$ are given by products of four $\epsilon_i$ which lie in the same plane:
\begin{equation}
(-1)^{\nu_{1,2,3}}=\prod_{n_i=1;n_{j\neq{i}}=0,1}\epsilon_{i=n_1n_2n_3}.
\label{eq:weakz2}
\end{equation}
$\nu_0=1$ indicates that the TI is topologically nontrivial. One can imagine slowly turning off the SOC, thereby closing the bulk gap. At the critical point between a topological and trivial insulator, the gap closes and a DNL forms, hosting the same topological indices as the TI. Moreover, Kim et al.\cite{Kim2015} showed that one can determine the number of DNLs intersecting any of the six invariant surfaces $S_{abcd}$ bounded by the four TRIM $a,b,c$ and $d$ by multiplying the $\epsilon_i$ at those four points:
\begin{equation}
(-1)^{N(S_{abcd})}=\epsilon_a\epsilon_b\epsilon_c\epsilon_d.
\label{eq:nDNL}
\end{equation}
If the product is $-1$, $N(S_{abcd})=1$ and an odd number of DNLs must pierce $S_{abcd}$. For the trivial case $N(S_{abcd})=0$ an even (including zero) number of DNLs pierce the surface.\par
\indent Using wavefunctions calculated with WIEN2k, we determine the parity eigenvalues at the eight TRIM in $\mathrm{TiRhAs}$ (a table is given in the supplementary material). The only TRIM point with $\epsilon_i=-1$ is $\Gamma$. Thus from Equations \ref{eq:strongz2} and \ref{eq:weakz2} we see that the topological indices for $\mathrm{TiRhAs}$ are $(\nu_0;\nu_1\nu_2\nu_3)=(1;000)$, and that $\mathrm{TiRhAs}$ is topologically robust. Additionally, the parities imply that the three invariant surfaces containing $\Gamma$, namely $k_x=0$, $k_y=0$, and $k_z=0$, are intersected by an odd number of DNLs, whereas the surfaces on the edge of the BZ are intersected by an even number of DNLs. This is completely consistent with our finding of the single DNL lying in the $k_x=0$ plane surrounding $\Gamma$.

\subsection{\label{surfacestate}Topological surface states}
\begin{figure}
\includegraphics{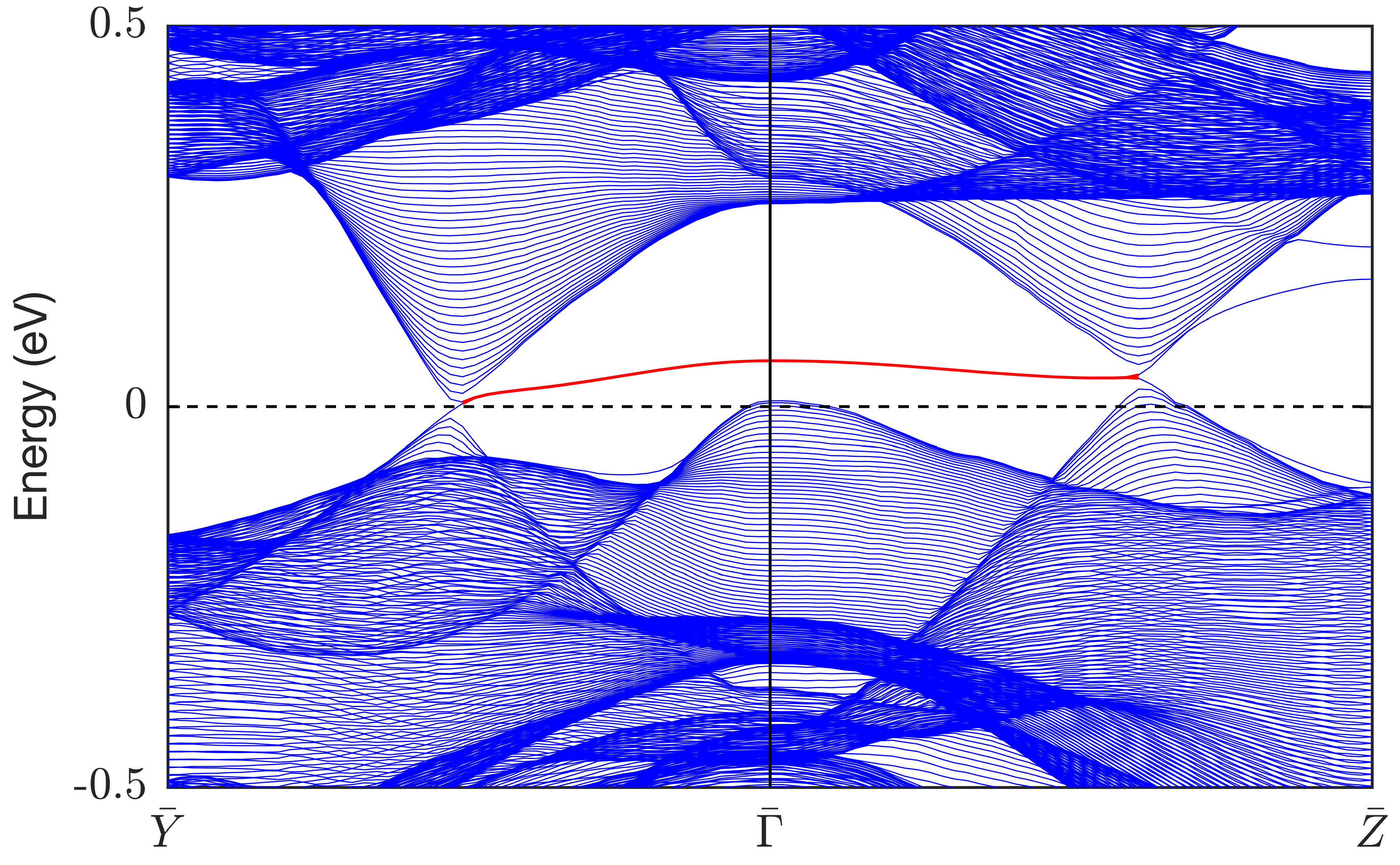}
\caption{\label{fig:surfstates}DFT-PBE tight-binding band structure (without SOC) for the $(100)$ surface plotted along the $\bar{Y}-\bar{\Gamma}-\bar{Z}$ direction, showing surface states (colored red) in the projected interior of the DNL. }
\end{figure}

Topologically robust nodal lines are predicted to host nearly flat, two dimensional drumhead surface states\cite{Chan2016} (see Supplementary Material for more rigorous justification). To study the surface states in $\mathrm{TiRhAs}$ we construct a tight-binding model from our DFT-PBE calculations using maximally localized Wannier functions (MLWFs)\cite{Marzari1997,Mostofi2014} as our basis states. We use 40 MLWFs derived from Ti \emph{d} and Rh \emph{d} bands around the Fermi level using a disentanglement procedure implemented in the open-source code Wannier90.\cite{Souza2001} Our model for a 100 unit cell thick slab in the $[100]$ direction is plotted for the $(100)$ surface in Figure \ref{fig:surfstates}. Large slabs are required to recover the bulk DNL in our calculations. Since our tight-binding model has two identical surfaces we see two completely degenerate surface states in the interior of the projected nodal line. The states have a slight dispersion due to the particle-hole asymmetry\cite{Yu2015}. \\
\indent We also construct a tight-binding model for the case where SOC is included. Because the effect of SOC is very slight in $\mathrm{TiRhAs}$, the qualitative band structure is very similar to Figure \ref{fig:surfstates}. However, since SOC introduces a continuous gap in the DNL the surface spectrum evolves from a nearly flat, drumhead state to a very shallow Dirac cone characteristic of TIs (see supplementary material).

\section{Conclusion}
\indent In summary we have performed extensive first-principles calculations on the previously synthesized compound $\mathrm{TiRhAs}$ and identify it as a new Dirac nodal line semimetal. The nodal line is topologically protected by both reflection symmetry and composite inversion and time-reversal symmetry; hence $\mathrm{TiRhAs}$ is particularly robust to local crystalline defects. We have performed a $\mathbf{k}\cdot \mathbf{p}$ analysis to determine the magnitude and tilting of band velocities along the DNL. We have calculated the $\mathbb{Z}_2$ invariants and have confirmed the presence of drumhead surface states. Moreover, the Fermi surface in $\mathrm{TiRhAs}$ is remarkably clean, and although SOC introduces gaps in the DNL, the effect is small. We therefore believe that further experimental studies on this compound should yield results consistent with our calculations.

\begin{acknowledgments}
The authors would like to thank S. Y. Park, S. Mack, and Q. S. Wu for useful discussions. This work is supported by the U.S. Department of Energy, Director, Office of Science, Office of Basic Energy Sciences, Materials Sciences and Engineering Division, under Contract No. DE-AC02-05CH11231, through the Theory FWP (KC2301) at Lawrence Berkeley National Laboratory (LBNL). This work is also supported by the Molecular Foundry through the DOE, Office of Basic Energy Sciences under the same contract number. Q. Y. was supported by the Center for Computational Design of Functional Layered Materials (CCDM), an Energy Frontier Research Center funded by the U.S. Department of Energy (DOE), Office of Science, Basic Energy Sciences (BES), under Award No. DE-SC0012575.  S. F. W. was supported under the National Defense Science and Engineering Graduate Fellowship (NDSEG). Calculations were performed on the Lawrencium cluster, operated by Lawrence Berkeley National Laboratory, and on the National Energy Research Scientific Computing Center (NERSC). 
\end{acknowledgments}

%






\begin{thebibliography}{48}%
\makeatletter
\providecommand \@ifxundefined [1]{%
 \@ifx{#1\undefined}
}%
\providecommand \@ifnum [1]{%
 \ifnum #1\expandafter \@firstoftwo
 \else \expandafter \@secondoftwo
 \fi
}%
\providecommand \@ifx [1]{%
 \ifx #1\expandafter \@firstoftwo
 \else \expandafter \@secondoftwo
 \fi
}%
\providecommand \natexlab [1]{#1}%
\providecommand \enquote  [1]{``#1''}%
\providecommand \bibnamefont  [1]{#1}%
\providecommand \bibfnamefont [1]{#1}%
\providecommand \citenamefont [1]{#1}%
\providecommand \href@noop [0]{\@secondoftwo}%
\providecommand \href [0]{\begingroup \@sanitize@url \@href}%
\providecommand \@href[1]{\@@startlink{#1}\@@href}%
\providecommand \@@href[1]{\endgroup#1\@@endlink}%
\providecommand \@sanitize@url [0]{\catcode `\\12\catcode `\$12\catcode
  `\&12\catcode `\#12\catcode `\^12\catcode `\_12\catcode `\%12\relax}%
\providecommand \@@startlink[1]{}%
\providecommand \@@endlink[0]{}%
\providecommand \url  [0]{\begingroup\@sanitize@url \@url }%
\providecommand \@url [1]{\endgroup\@href {#1}{\urlprefix }}%
\providecommand \urlprefix  [0]{URL }%
\providecommand \Eprint [0]{\href }%
\providecommand \doibase [0]{http://dx.doi.org/}%
\providecommand \selectlanguage [0]{\@gobble}%
\providecommand \bibinfo  [0]{\@secondoftwo}%
\providecommand \bibfield  [0]{\@secondoftwo}%
\providecommand \translation [1]{[#1]}%
\providecommand \BibitemOpen [0]{}%
\providecommand \bibitemStop [0]{}%
\providecommand \bibitemNoStop [0]{.\EOS\space}%
\providecommand \EOS [0]{\spacefactor3000\relax}%
\providecommand \BibitemShut  [1]{\csname bibitem#1\endcsname}%
\let\auto@bib@innerbib\@empty
\bibitem [{\citenamefont {Weng}\ \emph {et~al.}(2016)\citenamefont {Weng},
  \citenamefont {Dai},\ and\ \citenamefont {Fang}}]{Weng2016}%
  \BibitemOpen
  \bibfield  {author} {\bibinfo {author} {\bibfnamefont {H.}~\bibnamefont
  {Weng}}, \bibinfo {author} {\bibfnamefont {X.}~\bibnamefont {Dai}}, \ and\
  \bibinfo {author} {\bibfnamefont {Z.}~\bibnamefont {Fang}},\ }\href {\doibase
  10.1088/0953-8984/28/30/303001} {\bibfield  {journal} {\bibinfo  {journal}
  {Journal of Physics: Condensed Matter}\ }\textbf {\bibinfo {volume} {28}},\
  \bibinfo {pages} {303001} (\bibinfo {year} {2016})},\ \Eprint
  {http://arxiv.org/abs/1603.04744} {arXiv:1603.04744} \BibitemShut {NoStop}%
\bibitem [{\citenamefont {Murakami}(2007)}]{Murakami2007}%
  \BibitemOpen
  \bibfield  {author} {\bibinfo {author} {\bibfnamefont {S.}~\bibnamefont
  {Murakami}},\ }\href {\doibase 10.1088/1367-2630/9/9/356} {\bibfield
  {journal} {\bibinfo  {journal} {New Journal of Physics}\ }\textbf {\bibinfo
  {volume} {9}} (\bibinfo {year} {2007}),\ 10.1088/1367-2630/9/9/356},\ \Eprint
  {http://arxiv.org/abs/0710.0930} {arXiv:0710.0930} \BibitemShut {NoStop}%
\bibitem [{\citenamefont {Xu}\ \emph {et~al.}(2015)\citenamefont {Xu},
  \citenamefont {Belopolski}, \citenamefont {Alidoust}, \citenamefont
  {Neupane}, \citenamefont {Bian}, \citenamefont {Zhang}, \citenamefont
  {Sankar}, \citenamefont {Chang}, \citenamefont {Yuan}, \citenamefont {Lee},
  \citenamefont {Huang}, \citenamefont {Zheng}, \citenamefont {Ma},
  \citenamefont {Sanchez}, \citenamefont {Wang}, \citenamefont {Bansil},
  \citenamefont {Chou}, \citenamefont {Shibayev}, \citenamefont {Lin},
  \citenamefont {Jia},\ and\ \citenamefont {Hasan}}]{Xu2015}%
  \BibitemOpen
  \bibfield  {author} {\bibinfo {author} {\bibfnamefont {S.-Y.}\ \bibnamefont
  {Xu}}, \bibinfo {author} {\bibfnamefont {I.}~\bibnamefont {Belopolski}},
  \bibinfo {author} {\bibfnamefont {N.}~\bibnamefont {Alidoust}}, \bibinfo
  {author} {\bibfnamefont {M.}~\bibnamefont {Neupane}}, \bibinfo {author}
  {\bibfnamefont {G.}~\bibnamefont {Bian}}, \bibinfo {author} {\bibfnamefont
  {C.}~\bibnamefont {Zhang}}, \bibinfo {author} {\bibfnamefont
  {R.}~\bibnamefont {Sankar}}, \bibinfo {author} {\bibfnamefont
  {G.}~\bibnamefont {Chang}}, \bibinfo {author} {\bibfnamefont
  {Z.}~\bibnamefont {Yuan}}, \bibinfo {author} {\bibfnamefont {C.-C.}\
  \bibnamefont {Lee}}, \bibinfo {author} {\bibfnamefont {S.-M.}\ \bibnamefont
  {Huang}}, \bibinfo {author} {\bibfnamefont {H.}~\bibnamefont {Zheng}},
  \bibinfo {author} {\bibfnamefont {J.}~\bibnamefont {Ma}}, \bibinfo {author}
  {\bibfnamefont {D.~S.}\ \bibnamefont {Sanchez}}, \bibinfo {author}
  {\bibfnamefont {B.}~\bibnamefont {Wang}}, \bibinfo {author} {\bibfnamefont
  {A.}~\bibnamefont {Bansil}}, \bibinfo {author} {\bibfnamefont
  {F.}~\bibnamefont {Chou}}, \bibinfo {author} {\bibfnamefont {P.~P.}\
  \bibnamefont {Shibayev}}, \bibinfo {author} {\bibfnamefont {H.}~\bibnamefont
  {Lin}}, \bibinfo {author} {\bibfnamefont {S.}~\bibnamefont {Jia}}, \ and\
  \bibinfo {author} {\bibfnamefont {M.~Z.}\ \bibnamefont {Hasan}},\ }\href
  {\doibase 10.1126/science.aaa9297} {\bibfield  {journal} {\bibinfo  {journal}
  {Science}\ ,\ \bibinfo {pages} {science.aaa9297}} (\bibinfo {year} {2015})},\
  \Eprint {http://arxiv.org/abs/1502.03807} {arXiv:1502.03807} \BibitemShut
  {NoStop}%
\bibitem [{\citenamefont {Lv}\ \emph {et~al.}(2015)\citenamefont {Lv},
  \citenamefont {Weng}, \citenamefont {Fu}, \citenamefont {Wang}, \citenamefont
  {Miao}, \citenamefont {Ma}, \citenamefont {Richard}, \citenamefont {Huang},
  \citenamefont {Zhao}, \citenamefont {Chen}, \citenamefont {Fang},
  \citenamefont {Dai}, \citenamefont {Qian},\ and\ \citenamefont
  {Ding}}]{LV2015}%
  \BibitemOpen
  \bibfield  {author} {\bibinfo {author} {\bibfnamefont {B.~Q.}\ \bibnamefont
  {Lv}}, \bibinfo {author} {\bibfnamefont {H.~M.}\ \bibnamefont {Weng}},
  \bibinfo {author} {\bibfnamefont {B.~B.}\ \bibnamefont {Fu}}, \bibinfo
  {author} {\bibfnamefont {X.~P.}\ \bibnamefont {Wang}}, \bibinfo {author}
  {\bibfnamefont {H.}~\bibnamefont {Miao}}, \bibinfo {author} {\bibfnamefont
  {J.}~\bibnamefont {Ma}}, \bibinfo {author} {\bibfnamefont {P.}~\bibnamefont
  {Richard}}, \bibinfo {author} {\bibfnamefont {X.~C.}\ \bibnamefont {Huang}},
  \bibinfo {author} {\bibfnamefont {L.~X.}\ \bibnamefont {Zhao}}, \bibinfo
  {author} {\bibfnamefont {G.~F.}\ \bibnamefont {Chen}}, \bibinfo {author}
  {\bibfnamefont {Z.}~\bibnamefont {Fang}}, \bibinfo {author} {\bibfnamefont
  {X.}~\bibnamefont {Dai}}, \bibinfo {author} {\bibfnamefont {T.}~\bibnamefont
  {Qian}}, \ and\ \bibinfo {author} {\bibfnamefont {H.}~\bibnamefont {Ding}},\
  }\href {\doibase 10.1103/PhysRevX.5.031013} {\bibfield  {journal} {\bibinfo
  {journal} {Physical Review X}\ }\textbf {\bibinfo {volume} {5}},\ \bibinfo
  {pages} {031013} (\bibinfo {year} {2015})},\ \Eprint
  {http://arxiv.org/abs/1502.04684} {arXiv:1502.04684} \BibitemShut {NoStop}%
\bibitem [{\citenamefont {Wan}\ \emph {et~al.}(2011)\citenamefont {Wan},
  \citenamefont {Turner}, \citenamefont {Vishwanath},\ and\ \citenamefont
  {Savrasov}}]{Wan2011}%
  \BibitemOpen
  \bibfield  {author} {\bibinfo {author} {\bibfnamefont {X.}~\bibnamefont
  {Wan}}, \bibinfo {author} {\bibfnamefont {A.~M.}\ \bibnamefont {Turner}},
  \bibinfo {author} {\bibfnamefont {A.}~\bibnamefont {Vishwanath}}, \ and\
  \bibinfo {author} {\bibfnamefont {S.~Y.}\ \bibnamefont {Savrasov}},\ }\href
  {\doibase 10.1103/PhysRevB.83.205101} {\bibfield  {journal} {\bibinfo
  {journal} {Physical Review B - Condensed Matter and Materials Physics}\
  }\textbf {\bibinfo {volume} {83}},\ \bibinfo {pages} {1} (\bibinfo {year}
  {2011})},\ \Eprint {http://arxiv.org/abs/1007.0016} {arXiv:1007.0016}
  \BibitemShut {NoStop}%
\bibitem [{\citenamefont {Weng}\ \emph
  {et~al.}(2015{\natexlab{a}})\citenamefont {Weng}, \citenamefont {Fang},
  \citenamefont {Fang}, \citenamefont {{Andrei Bernevig}},\ and\ \citenamefont
  {Dai}}]{Weng2015a}%
  \BibitemOpen
  \bibfield  {author} {\bibinfo {author} {\bibfnamefont {H.}~\bibnamefont
  {Weng}}, \bibinfo {author} {\bibfnamefont {C.}~\bibnamefont {Fang}}, \bibinfo
  {author} {\bibfnamefont {Z.}~\bibnamefont {Fang}}, \bibinfo {author}
  {\bibfnamefont {B.}~\bibnamefont {{Andrei Bernevig}}}, \ and\ \bibinfo
  {author} {\bibfnamefont {X.}~\bibnamefont {Dai}},\ }\href {\doibase
  10.1103/PhysRevX.5.011029} {\bibfield  {journal} {\bibinfo  {journal}
  {Physical Review X}\ }\textbf {\bibinfo {volume} {5}},\ \bibinfo {pages} {1}
  (\bibinfo {year} {2015}{\natexlab{a}})},\ \Eprint
  {http://arxiv.org/abs/1501.00060} {arXiv:1501.00060} \BibitemShut {NoStop}%
\bibitem [{\citenamefont {Liu}\ \emph {et~al.}(2014{\natexlab{a}})\citenamefont
  {Liu}, \citenamefont {Jiang}, \citenamefont {Zhou}, \citenamefont {Wang},
  \citenamefont {Zhang}, \citenamefont {Weng}, \citenamefont {Prabhakaran},
  \citenamefont {Mo}, \citenamefont {Peng}, \citenamefont {Dudin},
  \citenamefont {Kim}, \citenamefont {Hoesch}, \citenamefont {Fang},
  \citenamefont {Dai}, \citenamefont {Shen}, \citenamefont {Feng},
  \citenamefont {Hussain},\ and\ \citenamefont {Chen}}]{Liu2014a}%
  \BibitemOpen
  \bibfield  {author} {\bibinfo {author} {\bibfnamefont {Z.~K.}\ \bibnamefont
  {Liu}}, \bibinfo {author} {\bibfnamefont {J.}~\bibnamefont {Jiang}}, \bibinfo
  {author} {\bibfnamefont {B.}~\bibnamefont {Zhou}}, \bibinfo {author}
  {\bibfnamefont {Z.~J.}\ \bibnamefont {Wang}}, \bibinfo {author}
  {\bibfnamefont {Y.}~\bibnamefont {Zhang}}, \bibinfo {author} {\bibfnamefont
  {H.~M.}\ \bibnamefont {Weng}}, \bibinfo {author} {\bibfnamefont
  {D.}~\bibnamefont {Prabhakaran}}, \bibinfo {author} {\bibfnamefont {S.-k.}\
  \bibnamefont {Mo}}, \bibinfo {author} {\bibfnamefont {H.}~\bibnamefont
  {Peng}}, \bibinfo {author} {\bibfnamefont {P.}~\bibnamefont {Dudin}},
  \bibinfo {author} {\bibfnamefont {T.}~\bibnamefont {Kim}}, \bibinfo {author}
  {\bibfnamefont {M.}~\bibnamefont {Hoesch}}, \bibinfo {author} {\bibfnamefont
  {Z.}~\bibnamefont {Fang}}, \bibinfo {author} {\bibfnamefont {X.}~\bibnamefont
  {Dai}}, \bibinfo {author} {\bibfnamefont {Z.~X.}\ \bibnamefont {Shen}},
  \bibinfo {author} {\bibfnamefont {D.~L.}\ \bibnamefont {Feng}}, \bibinfo
  {author} {\bibfnamefont {Z.}~\bibnamefont {Hussain}}, \ and\ \bibinfo
  {author} {\bibfnamefont {Y.~L.}\ \bibnamefont {Chen}},\ }\href {\doibase
  10.1038/nmat3990} {\bibfield  {journal} {\bibinfo  {journal} {Nature
  Materials}\ }\textbf {\bibinfo {volume} {13}},\ \bibinfo {pages} {677}
  (\bibinfo {year} {2014}{\natexlab{a}})},\ \Eprint
  {http://arxiv.org/abs/1310.0391} {arXiv:1310.0391} \BibitemShut {NoStop}%
\bibitem [{\citenamefont {Liu}\ \emph {et~al.}(2014{\natexlab{b}})\citenamefont
  {Liu}, \citenamefont {Zhou}, \citenamefont {Zhang}, \citenamefont {Wang},
  \citenamefont {Weng}, \citenamefont {Prabhakaran}, \citenamefont {Mo},
  \citenamefont {Shen}, \citenamefont {Fang}, \citenamefont {Dai},
  \citenamefont {Hussain},\ and\ \citenamefont {Chen}}]{Liu2014}%
  \BibitemOpen
  \bibfield  {author} {\bibinfo {author} {\bibfnamefont {Z.~K.}\ \bibnamefont
  {Liu}}, \bibinfo {author} {\bibfnamefont {B.}~\bibnamefont {Zhou}}, \bibinfo
  {author} {\bibfnamefont {Y.}~\bibnamefont {Zhang}}, \bibinfo {author}
  {\bibfnamefont {Z.~J.}\ \bibnamefont {Wang}}, \bibinfo {author}
  {\bibfnamefont {H.~M.}\ \bibnamefont {Weng}}, \bibinfo {author}
  {\bibfnamefont {D.}~\bibnamefont {Prabhakaran}}, \bibinfo {author}
  {\bibfnamefont {S.}~\bibnamefont {Mo}}, \bibinfo {author} {\bibfnamefont
  {Z.~X.}\ \bibnamefont {Shen}}, \bibinfo {author} {\bibfnamefont
  {Z.}~\bibnamefont {Fang}}, \bibinfo {author} {\bibfnamefont {X.}~\bibnamefont
  {Dai}}, \bibinfo {author} {\bibfnamefont {Z.}~\bibnamefont {Hussain}}, \ and\
  \bibinfo {author} {\bibfnamefont {Y.~L.}\ \bibnamefont {Chen}},\ }\href
  {\doibase 10.1126/science.1245085} {\bibfield  {journal} {\bibinfo  {journal}
  {Science}\ }\textbf {\bibinfo {volume} {343}},\ \bibinfo {pages} {864}
  (\bibinfo {year} {2014}{\natexlab{b}})}\BibitemShut {NoStop}%
\bibitem [{\citenamefont {Wang}\ \emph {et~al.}(2013)\citenamefont {Wang},
  \citenamefont {Weng}, \citenamefont {Wu}, \citenamefont {Dai},\ and\
  \citenamefont {Fang}}]{Wang2013}%
  \BibitemOpen
  \bibfield  {author} {\bibinfo {author} {\bibfnamefont {Z.}~\bibnamefont
  {Wang}}, \bibinfo {author} {\bibfnamefont {H.}~\bibnamefont {Weng}}, \bibinfo
  {author} {\bibfnamefont {Q.}~\bibnamefont {Wu}}, \bibinfo {author}
  {\bibfnamefont {X.}~\bibnamefont {Dai}}, \ and\ \bibinfo {author}
  {\bibfnamefont {Z.}~\bibnamefont {Fang}},\ }\href {\doibase
  10.1103/PhysRevB.88.125427} {\bibfield  {journal} {\bibinfo  {journal}
  {Physical Review B - Condensed Matter and Materials Physics}\ }\textbf
  {\bibinfo {volume} {88}},\ \bibinfo {pages} {1} (\bibinfo {year} {2013})},\
  \Eprint {http://arxiv.org/abs/arXiv:1305.6780v1} {arXiv:arXiv:1305.6780v1}
  \BibitemShut {NoStop}%
\bibitem [{\citenamefont {Hu}\ \emph {et~al.}(2016)\citenamefont {Hu},
  \citenamefont {Tang}, \citenamefont {Liu}, \citenamefont {Liu}, \citenamefont
  {Zhu}, \citenamefont {Graf}, \citenamefont {Myhro}, \citenamefont {Tran},
  \citenamefont {Lau}, \citenamefont {Wei},\ and\ \citenamefont
  {Mao}}]{Hu2016}%
  \BibitemOpen
  \bibfield  {author} {\bibinfo {author} {\bibfnamefont {J.}~\bibnamefont
  {Hu}}, \bibinfo {author} {\bibfnamefont {Z.}~\bibnamefont {Tang}}, \bibinfo
  {author} {\bibfnamefont {J.}~\bibnamefont {Liu}}, \bibinfo {author}
  {\bibfnamefont {X.}~\bibnamefont {Liu}}, \bibinfo {author} {\bibfnamefont
  {Y.}~\bibnamefont {Zhu}}, \bibinfo {author} {\bibfnamefont {D.}~\bibnamefont
  {Graf}}, \bibinfo {author} {\bibfnamefont {K.}~\bibnamefont {Myhro}},
  \bibinfo {author} {\bibfnamefont {S.}~\bibnamefont {Tran}}, \bibinfo {author}
  {\bibfnamefont {C.~N.}\ \bibnamefont {Lau}}, \bibinfo {author} {\bibfnamefont
  {J.}~\bibnamefont {Wei}}, \ and\ \bibinfo {author} {\bibfnamefont
  {Z.}~\bibnamefont {Mao}},\ }\href {\doibase 10.1103/PhysRevLett.117.016602}
  {\bibfield  {journal} {\bibinfo  {journal} {Physical Review Letters}\
  }\textbf {\bibinfo {volume} {117}},\ \bibinfo {pages} {1} (\bibinfo {year}
  {2016})},\ \Eprint {http://arxiv.org/abs/1604.06860} {arXiv:1604.06860}
  \BibitemShut {NoStop}%
\bibitem [{\citenamefont {Bian}\ \emph
  {et~al.}(2016{\natexlab{a}})\citenamefont {Bian}, \citenamefont {Chang},
  \citenamefont {Sankar}, \citenamefont {Xu}, \citenamefont {Zheng},
  \citenamefont {Neupert}, \citenamefont {Chiu}, \citenamefont {Huang},
  \citenamefont {Chang}, \citenamefont {Belopolski}, \citenamefont {Sanchez},
  \citenamefont {Neupane}, \citenamefont {Alidoust}, \citenamefont {Liu},
  \citenamefont {Wang}, \citenamefont {Lee}, \citenamefont {Jeng},
  \citenamefont {Zhang}, \citenamefont {Yuan}, \citenamefont {Jia},
  \citenamefont {Bansil}, \citenamefont {Chou}, \citenamefont {Lin},\ and\
  \citenamefont {Hasan}}]{Bian2016a}%
  \BibitemOpen
  \bibfield  {author} {\bibinfo {author} {\bibfnamefont {G.}~\bibnamefont
  {Bian}}, \bibinfo {author} {\bibfnamefont {T.-R.}\ \bibnamefont {Chang}},
  \bibinfo {author} {\bibfnamefont {R.}~\bibnamefont {Sankar}}, \bibinfo
  {author} {\bibfnamefont {S.-Y.}\ \bibnamefont {Xu}}, \bibinfo {author}
  {\bibfnamefont {H.}~\bibnamefont {Zheng}}, \bibinfo {author} {\bibfnamefont
  {T.}~\bibnamefont {Neupert}}, \bibinfo {author} {\bibfnamefont {C.-K.}\
  \bibnamefont {Chiu}}, \bibinfo {author} {\bibfnamefont {S.-M.}\ \bibnamefont
  {Huang}}, \bibinfo {author} {\bibfnamefont {G.}~\bibnamefont {Chang}},
  \bibinfo {author} {\bibfnamefont {I.}~\bibnamefont {Belopolski}}, \bibinfo
  {author} {\bibfnamefont {D.~S.}\ \bibnamefont {Sanchez}}, \bibinfo {author}
  {\bibfnamefont {M.}~\bibnamefont {Neupane}}, \bibinfo {author} {\bibfnamefont
  {N.}~\bibnamefont {Alidoust}}, \bibinfo {author} {\bibfnamefont
  {C.}~\bibnamefont {Liu}}, \bibinfo {author} {\bibfnamefont {B.}~\bibnamefont
  {Wang}}, \bibinfo {author} {\bibfnamefont {C.-C.}\ \bibnamefont {Lee}},
  \bibinfo {author} {\bibfnamefont {H.-T.}\ \bibnamefont {Jeng}}, \bibinfo
  {author} {\bibfnamefont {C.}~\bibnamefont {Zhang}}, \bibinfo {author}
  {\bibfnamefont {Z.}~\bibnamefont {Yuan}}, \bibinfo {author} {\bibfnamefont
  {S.}~\bibnamefont {Jia}}, \bibinfo {author} {\bibfnamefont {A.}~\bibnamefont
  {Bansil}}, \bibinfo {author} {\bibfnamefont {F.}~\bibnamefont {Chou}},
  \bibinfo {author} {\bibfnamefont {H.}~\bibnamefont {Lin}}, \ and\ \bibinfo
  {author} {\bibfnamefont {M.~Z.}\ \bibnamefont {Hasan}},\ }\href {\doibase
  10.1038/ncomms10556} {\bibfield  {journal} {\bibinfo  {journal} {Nature
  communications}\ }\textbf {\bibinfo {volume} {7}},\ \bibinfo {pages} {10556}
  (\bibinfo {year} {2016}{\natexlab{a}})},\ \Eprint
  {http://arxiv.org/abs/1505.03069} {arXiv:1505.03069} \BibitemShut {NoStop}%
\bibitem [{\citenamefont {Neupane}\ \emph {et~al.}(2016)\citenamefont
  {Neupane}, \citenamefont {Belopolski}, \citenamefont {Hosen}, \citenamefont
  {Sanchez}, \citenamefont {Sankar}, \citenamefont {Szlawska}, \citenamefont
  {Xu}, \citenamefont {Dimitri}, \citenamefont {Dhakal}, \citenamefont
  {Maldonado}, \citenamefont {Oppeneer}, \citenamefont {Kaczorowski},
  \citenamefont {Chou}, \citenamefont {Hasan},\ and\ \citenamefont
  {Durakiewicz}}]{Neupane2016}%
  \BibitemOpen
  \bibfield  {author} {\bibinfo {author} {\bibfnamefont {M.}~\bibnamefont
  {Neupane}}, \bibinfo {author} {\bibfnamefont {I.}~\bibnamefont {Belopolski}},
  \bibinfo {author} {\bibfnamefont {M.~M.}\ \bibnamefont {Hosen}}, \bibinfo
  {author} {\bibfnamefont {D.~S.}\ \bibnamefont {Sanchez}}, \bibinfo {author}
  {\bibfnamefont {R.}~\bibnamefont {Sankar}}, \bibinfo {author} {\bibfnamefont
  {M.}~\bibnamefont {Szlawska}}, \bibinfo {author} {\bibfnamefont {S.~Y.}\
  \bibnamefont {Xu}}, \bibinfo {author} {\bibfnamefont {K.}~\bibnamefont
  {Dimitri}}, \bibinfo {author} {\bibfnamefont {N.}~\bibnamefont {Dhakal}},
  \bibinfo {author} {\bibfnamefont {P.}~\bibnamefont {Maldonado}}, \bibinfo
  {author} {\bibfnamefont {P.~M.}\ \bibnamefont {Oppeneer}}, \bibinfo {author}
  {\bibfnamefont {D.}~\bibnamefont {Kaczorowski}}, \bibinfo {author}
  {\bibfnamefont {F.}~\bibnamefont {Chou}}, \bibinfo {author} {\bibfnamefont
  {M.~Z.}\ \bibnamefont {Hasan}}, \ and\ \bibinfo {author} {\bibfnamefont
  {T.}~\bibnamefont {Durakiewicz}},\ }\href {\doibase
  10.1103/PhysRevB.93.201104} {\bibfield  {journal} {\bibinfo  {journal}
  {Physical Review B}\ }\textbf {\bibinfo {volume} {93}},\ \bibinfo {pages} {1}
  (\bibinfo {year} {2016})},\ \Eprint {http://arxiv.org/abs/1604.00720}
  {arXiv:1604.00720} \BibitemShut {NoStop}%
\bibitem [{\citenamefont {Yu}\ \emph {et~al.}(2015)\citenamefont {Yu},
  \citenamefont {Weng}, \citenamefont {Fang}, \citenamefont {Dai},\ and\
  \citenamefont {Hu}}]{Yu2015}%
  \BibitemOpen
  \bibfield  {author} {\bibinfo {author} {\bibfnamefont {R.}~\bibnamefont
  {Yu}}, \bibinfo {author} {\bibfnamefont {H.}~\bibnamefont {Weng}}, \bibinfo
  {author} {\bibfnamefont {Z.}~\bibnamefont {Fang}}, \bibinfo {author}
  {\bibfnamefont {X.}~\bibnamefont {Dai}}, \ and\ \bibinfo {author}
  {\bibfnamefont {X.}~\bibnamefont {Hu}},\ }\href {\doibase
  10.1103/PhysRevLett.115.036807} {\bibfield  {journal} {\bibinfo  {journal}
  {Physical Review Letters}\ }\textbf {\bibinfo {volume} {115}},\ \bibinfo
  {pages} {3} (\bibinfo {year} {2015})},\ \Eprint
  {http://arxiv.org/abs/1504.04577} {arXiv:1504.04577} \BibitemShut {NoStop}%
\bibitem [{\citenamefont {Liang}\ \emph {et~al.}(2014)\citenamefont {Liang},
  \citenamefont {Gibson}, \citenamefont {Ali}, \citenamefont {Liu},
  \citenamefont {Cava},\ and\ \citenamefont {Ong}}]{Liang2014}%
  \BibitemOpen
  \bibfield  {author} {\bibinfo {author} {\bibfnamefont {T.}~\bibnamefont
  {Liang}}, \bibinfo {author} {\bibfnamefont {Q.}~\bibnamefont {Gibson}},
  \bibinfo {author} {\bibfnamefont {M.~N.}\ \bibnamefont {Ali}}, \bibinfo
  {author} {\bibfnamefont {M.}~\bibnamefont {Liu}}, \bibinfo {author}
  {\bibfnamefont {R.~J.}\ \bibnamefont {Cava}}, \ and\ \bibinfo {author}
  {\bibfnamefont {N.~P.}\ \bibnamefont {Ong}},\ }\href {\doibase
  10.1038/nmat4143} {\bibfield  {journal} {\bibinfo  {journal} {Nature
  Materials}\ }\textbf {\bibinfo {volume} {14}},\ \bibinfo {pages} {280}
  (\bibinfo {year} {2014})},\ \Eprint {http://arxiv.org/abs/1404.7794}
  {arXiv:1404.7794} \BibitemShut {NoStop}%
\bibitem [{\citenamefont {Parameswaran}\ \emph {et~al.}(2014)\citenamefont
  {Parameswaran}, \citenamefont {Grover}, \citenamefont {Abanin}, \citenamefont
  {Pesin},\ and\ \citenamefont {Vishwanath}}]{Parameswaran2014}%
  \BibitemOpen
  \bibfield  {author} {\bibinfo {author} {\bibfnamefont {S.~A.}\ \bibnamefont
  {Parameswaran}}, \bibinfo {author} {\bibfnamefont {T.}~\bibnamefont
  {Grover}}, \bibinfo {author} {\bibfnamefont {D.~A.}\ \bibnamefont {Abanin}},
  \bibinfo {author} {\bibfnamefont {D.~A.}\ \bibnamefont {Pesin}}, \ and\
  \bibinfo {author} {\bibfnamefont {A.}~\bibnamefont {Vishwanath}},\ }\href
  {\doibase 10.1103/PhysRevX.4.031035} {\bibfield  {journal} {\bibinfo
  {journal} {Physical Review X}\ }\textbf {\bibinfo {volume} {4}},\ \bibinfo
  {pages} {1} (\bibinfo {year} {2014})},\ \Eprint
  {http://arxiv.org/abs/1306.1234} {arXiv:1306.1234} \BibitemShut {NoStop}%
\bibitem [{\citenamefont {Bian}\ \emph
  {et~al.}(2016{\natexlab{b}})\citenamefont {Bian}, \citenamefont {Chang},
  \citenamefont {Zheng}, \citenamefont {Velury}, \citenamefont {Xu},
  \citenamefont {Neupert}, \citenamefont {Chiu}, \citenamefont {Huang},
  \citenamefont {Sanchez}, \citenamefont {Belopolski}, \citenamefont
  {Alidoust}, \citenamefont {Chen}, \citenamefont {Chang}, \citenamefont
  {Bansil}, \citenamefont {Jeng}, \citenamefont {Lin},\ and\ \citenamefont
  {Hasan}}]{Bian2016b}%
  \BibitemOpen
  \bibfield  {author} {\bibinfo {author} {\bibfnamefont {G.}~\bibnamefont
  {Bian}}, \bibinfo {author} {\bibfnamefont {T.~R.}\ \bibnamefont {Chang}},
  \bibinfo {author} {\bibfnamefont {H.}~\bibnamefont {Zheng}}, \bibinfo
  {author} {\bibfnamefont {S.}~\bibnamefont {Velury}}, \bibinfo {author}
  {\bibfnamefont {S.~Y.}\ \bibnamefont {Xu}}, \bibinfo {author} {\bibfnamefont
  {T.}~\bibnamefont {Neupert}}, \bibinfo {author} {\bibfnamefont {C.~K.}\
  \bibnamefont {Chiu}}, \bibinfo {author} {\bibfnamefont {S.~M.}\ \bibnamefont
  {Huang}}, \bibinfo {author} {\bibfnamefont {D.~S.}\ \bibnamefont {Sanchez}},
  \bibinfo {author} {\bibfnamefont {I.}~\bibnamefont {Belopolski}}, \bibinfo
  {author} {\bibfnamefont {N.}~\bibnamefont {Alidoust}}, \bibinfo {author}
  {\bibfnamefont {P.~J.}\ \bibnamefont {Chen}}, \bibinfo {author}
  {\bibfnamefont {G.}~\bibnamefont {Chang}}, \bibinfo {author} {\bibfnamefont
  {A.}~\bibnamefont {Bansil}}, \bibinfo {author} {\bibfnamefont {H.~T.}\
  \bibnamefont {Jeng}}, \bibinfo {author} {\bibfnamefont {H.}~\bibnamefont
  {Lin}}, \ and\ \bibinfo {author} {\bibfnamefont {M.~Z.}\ \bibnamefont
  {Hasan}},\ }\href {\doibase 10.1103/PhysRevB.93.121113} {\bibfield  {journal}
  {\bibinfo  {journal} {Physical Review B}\ }\textbf {\bibinfo {volume} {93}},\
  \bibinfo {pages} {2} (\bibinfo {year} {2016}{\natexlab{b}})},\ \Eprint
  {http://arxiv.org/abs/1508.07521} {arXiv:1508.07521} \BibitemShut {NoStop}%
\bibitem [{\citenamefont {Huh}\ \emph {et~al.}(2016)\citenamefont {Huh},
  \citenamefont {Moon},\ and\ \citenamefont {Kim}}]{Huh2016}%
  \BibitemOpen
  \bibfield  {author} {\bibinfo {author} {\bibfnamefont {Y.}~\bibnamefont
  {Huh}}, \bibinfo {author} {\bibfnamefont {E.~G.}\ \bibnamefont {Moon}}, \
  and\ \bibinfo {author} {\bibfnamefont {Y.~B.}\ \bibnamefont {Kim}},\ }\href
  {\doibase 10.1103/PhysRevB.93.035138} {\bibfield  {journal} {\bibinfo
  {journal} {Physical Review B}\ }\textbf {\bibinfo {volume} {93}} (\bibinfo
  {year} {2016}),\ 10.1103/PhysRevB.93.035138},\ \Eprint
  {http://arxiv.org/abs/1506.05105} {arXiv:1506.05105} \BibitemShut {NoStop}%
\bibitem [{\citenamefont {Chan}\ \emph
  {et~al.}(2016{\natexlab{a}})\citenamefont {Chan}, \citenamefont {Chiu},
  \citenamefont {Chou},\ and\ \citenamefont {Schnyder}}]{Chan2016}%
  \BibitemOpen
  \bibfield  {author} {\bibinfo {author} {\bibfnamefont {Y.~H.}\ \bibnamefont
  {Chan}}, \bibinfo {author} {\bibfnamefont {C.~K.}\ \bibnamefont {Chiu}},
  \bibinfo {author} {\bibfnamefont {M.~Y.}\ \bibnamefont {Chou}}, \ and\
  \bibinfo {author} {\bibfnamefont {A.~P.}\ \bibnamefont {Schnyder}},\ }\href
  {\doibase 10.1103/PhysRevB.93.205132} {\bibfield  {journal} {\bibinfo
  {journal} {Physical Review B}\ }\textbf {\bibinfo {volume} {93}},\ \bibinfo
  {pages} {1} (\bibinfo {year} {2016}{\natexlab{a}})},\ \Eprint
  {http://arxiv.org/abs/1510.02759} {arXiv:1510.02759} \BibitemShut {NoStop}%
\bibitem [{\citenamefont {Burkov}\ \emph {et~al.}(2011)\citenamefont {Burkov},
  \citenamefont {Hook},\ and\ \citenamefont {Balents}}]{Burkov2011}%
  \BibitemOpen
  \bibfield  {author} {\bibinfo {author} {\bibfnamefont {A.~A.}\ \bibnamefont
  {Burkov}}, \bibinfo {author} {\bibfnamefont {M.~D.}\ \bibnamefont {Hook}}, \
  and\ \bibinfo {author} {\bibfnamefont {L.}~\bibnamefont {Balents}},\ }\href
  {\doibase 10.1103/PhysRevB.84.235126} {\bibfield  {journal} {\bibinfo
  {journal} {Physical Review B - Condensed Matter and Materials Physics}\
  }\textbf {\bibinfo {volume} {84}},\ \bibinfo {pages} {1} (\bibinfo {year}
  {2011})},\ \Eprint {http://arxiv.org/abs/1110.1089} {arXiv:1110.1089}
  \BibitemShut {NoStop}%
\bibitem [{\citenamefont {Fang}\ \emph {et~al.}(2016)\citenamefont {Fang},
  \citenamefont {Weng}, \citenamefont {Dai},\ and\ \citenamefont
  {Fang}}]{Fang2016}%
  \BibitemOpen
  \bibfield  {author} {\bibinfo {author} {\bibfnamefont {C.}~\bibnamefont
  {Fang}}, \bibinfo {author} {\bibfnamefont {H.}~\bibnamefont {Weng}}, \bibinfo
  {author} {\bibfnamefont {X.}~\bibnamefont {Dai}}, \ and\ \bibinfo {author}
  {\bibfnamefont {Z.}~\bibnamefont {Fang}},\ }\href {\doibase
  10.1088/1674-1056/25/11/117106} {\bibfield  {journal} {\bibinfo  {journal}
  {Chinese Physics B}\ }\textbf {\bibinfo {volume} {25}},\ \bibinfo {pages}
  {117106} (\bibinfo {year} {2016})},\ \Eprint
  {http://arxiv.org/abs/1609.05414} {arXiv:1609.05414} \BibitemShut {NoStop}%
\bibitem [{\citenamefont {Fang}\ \emph {et~al.}(2015)\citenamefont {Fang},
  \citenamefont {Chen}, \citenamefont {Kee},\ and\ \citenamefont
  {Fu}}]{Fang2015}%
  \BibitemOpen
  \bibfield  {author} {\bibinfo {author} {\bibfnamefont {C.}~\bibnamefont
  {Fang}}, \bibinfo {author} {\bibfnamefont {Y.}~\bibnamefont {Chen}}, \bibinfo
  {author} {\bibfnamefont {H.~Y.}\ \bibnamefont {Kee}}, \ and\ \bibinfo
  {author} {\bibfnamefont {L.}~\bibnamefont {Fu}},\ }\href {\doibase
  10.1103/PhysRevB.92.081201} {\bibfield  {journal} {\bibinfo  {journal}
  {Physical Review B - Condensed Matter and Materials Physics}\ }\textbf
  {\bibinfo {volume} {92}},\ \bibinfo {pages} {1} (\bibinfo {year} {2015})},\
  \Eprint {http://arxiv.org/abs/1506.03449} {arXiv:1506.03449} \BibitemShut
  {NoStop}%
\bibitem [{\citenamefont {Chen}\ \emph {et~al.}(2017)\citenamefont {Chen},
  \citenamefont {Xu}, \citenamefont {Jiang}, \citenamefont {Wu}, \citenamefont
  {Qi}, \citenamefont {Yang}, \citenamefont {Wang}, \citenamefont {Sun},
  \citenamefont {Schr{\"{o}}ter}, \citenamefont {Yang}, \citenamefont {Schoop},
  \citenamefont {Lv}, \citenamefont {Zhou}, \citenamefont {Chen}, \citenamefont
  {Yao}, \citenamefont {Lu}, \citenamefont {Chen}, \citenamefont {Felser},
  \citenamefont {Yan}, \citenamefont {Liu},\ and\ \citenamefont
  {Chen}}]{Chen2017}%
  \BibitemOpen
  \bibfield  {author} {\bibinfo {author} {\bibfnamefont {C.}~\bibnamefont
  {Chen}}, \bibinfo {author} {\bibfnamefont {X.}~\bibnamefont {Xu}}, \bibinfo
  {author} {\bibfnamefont {J.}~\bibnamefont {Jiang}}, \bibinfo {author}
  {\bibfnamefont {S.~C.}\ \bibnamefont {Wu}}, \bibinfo {author} {\bibfnamefont
  {Y.~P.}\ \bibnamefont {Qi}}, \bibinfo {author} {\bibfnamefont {L.~X.}\
  \bibnamefont {Yang}}, \bibinfo {author} {\bibfnamefont {M.~X.}\ \bibnamefont
  {Wang}}, \bibinfo {author} {\bibfnamefont {Y.}~\bibnamefont {Sun}}, \bibinfo
  {author} {\bibfnamefont {N.~B.}\ \bibnamefont {Schr{\"{o}}ter}}, \bibinfo
  {author} {\bibfnamefont {H.~F.}\ \bibnamefont {Yang}}, \bibinfo {author}
  {\bibfnamefont {L.~M.}\ \bibnamefont {Schoop}}, \bibinfo {author}
  {\bibfnamefont {Y.~Y.}\ \bibnamefont {Lv}}, \bibinfo {author} {\bibfnamefont
  {J.}~\bibnamefont {Zhou}}, \bibinfo {author} {\bibfnamefont {Y.~B.}\
  \bibnamefont {Chen}}, \bibinfo {author} {\bibfnamefont {S.~H.}\ \bibnamefont
  {Yao}}, \bibinfo {author} {\bibfnamefont {M.~H.}\ \bibnamefont {Lu}},
  \bibinfo {author} {\bibfnamefont {Y.~F.}\ \bibnamefont {Chen}}, \bibinfo
  {author} {\bibfnamefont {C.}~\bibnamefont {Felser}}, \bibinfo {author}
  {\bibfnamefont {B.~H.}\ \bibnamefont {Yan}}, \bibinfo {author} {\bibfnamefont
  {Z.~K.}\ \bibnamefont {Liu}}, \ and\ \bibinfo {author} {\bibfnamefont
  {Y.~L.}\ \bibnamefont {Chen}},\ }\href {\doibase 10.1103/PhysRevB.95.125126}
  {\bibfield  {journal} {\bibinfo  {journal} {Physical Review B}\ }\textbf
  {\bibinfo {volume} {95}},\ \bibinfo {pages} {1} (\bibinfo {year} {2017})},\
  \Eprint {http://arxiv.org/abs/1701.06888} {arXiv:1701.06888} \BibitemShut
  {NoStop}%
\bibitem [{\citenamefont {Roy-Montreuil}\ \emph {et~al.}(1984)\citenamefont
  {Roy-Montreuil}, \citenamefont {Chaudouet}, \citenamefont {Boursier},
  \citenamefont {Senateur},\ and\ \citenamefont
  {Fruchart}}]{Roy-Montreuil1984}%
  \BibitemOpen
  \bibfield  {author} {\bibinfo {author} {\bibfnamefont {J.}~\bibnamefont
  {Roy-Montreuil}}, \bibinfo {author} {\bibfnamefont {P.}~\bibnamefont
  {Chaudouet}}, \bibinfo {author} {\bibfnamefont {D.}~\bibnamefont {Boursier}},
  \bibinfo {author} {\bibfnamefont {J.}~\bibnamefont {Senateur}}, \ and\
  \bibinfo {author} {\bibfnamefont {R.}~\bibnamefont {Fruchart}},\ }\href@noop
  {} {\bibfield  {journal} {\bibinfo  {journal} {Annales de Chimie}\ }\textbf
  {\bibinfo {volume} {9}} (\bibinfo {year} {1984})}\BibitemShut {NoStop}%
\bibitem [{\citenamefont {Luttinger}\ and\ \citenamefont
  {Ward}(1960)}]{Luttinger1960}%
  \BibitemOpen
  \bibfield  {author} {\bibinfo {author} {\bibfnamefont {J.~M.}\ \bibnamefont
  {Luttinger}}\ and\ \bibinfo {author} {\bibfnamefont {J.~C.}\ \bibnamefont
  {Ward}},\ }\href {\doibase 10.1017/CBO9781107415324.004} {\bibfield
  {journal} {\bibinfo  {journal} {Physical Review}\ }\textbf {\bibinfo {volume}
  {118}},\ \bibinfo {pages} {1417} (\bibinfo {year} {1960})},\ \Eprint
  {http://arxiv.org/abs/arXiv:1011.1669v3} {arXiv:arXiv:1011.1669v3}
  \BibitemShut {NoStop}%
\bibitem [{\citenamefont {Kresse}\ and\ \citenamefont
  {Furthm{\"{u}}ller}(1996)}]{Kresse1996}%
  \BibitemOpen
  \bibfield  {author} {\bibinfo {author} {\bibfnamefont {G.}~\bibnamefont
  {Kresse}}\ and\ \bibinfo {author} {\bibfnamefont {J.}~\bibnamefont
  {Furthm{\"{u}}ller}},\ }\href {\doibase 10.1103/PhysRevB.54.11169} {\bibfield
   {journal} {\bibinfo  {journal} {Physical Review B}\ }\textbf {\bibinfo
  {volume} {54}},\ \bibinfo {pages} {11169} (\bibinfo {year} {1996})},\ \Eprint
  {http://arxiv.org/abs/0927-0256(96)00008} {arXiv:0927-0256(96)00008
  [10.1016]} \BibitemShut {NoStop}%
\bibitem [{\citenamefont {Perdew}\ \emph {et~al.}(1996)\citenamefont {Perdew},
  \citenamefont {Burke},\ and\ \citenamefont {Ernzerhof}}]{Perdew1996}%
  \BibitemOpen
  \bibfield  {author} {\bibinfo {author} {\bibfnamefont {J.~P.}\ \bibnamefont
  {Perdew}}, \bibinfo {author} {\bibfnamefont {K.}~\bibnamefont {Burke}}, \
  and\ \bibinfo {author} {\bibfnamefont {M.}~\bibnamefont {Ernzerhof}},\ }\href
  {\doibase 10.1103/PhysRevLett.77.3865} {\bibfield  {journal} {\bibinfo
  {journal} {Physical Review Letters}\ }\textbf {\bibinfo {volume} {77}},\
  \bibinfo {pages} {3865} (\bibinfo {year} {1996})},\ \Eprint
  {http://arxiv.org/abs/0927-0256(96)00008} {arXiv:0927-0256(96)00008
  [10.1016]} \BibitemShut {NoStop}%
\bibitem [{\citenamefont {Bl{\"{o}}chl}(1994)}]{Blochl1994}%
  \BibitemOpen
  \bibfield  {author} {\bibinfo {author} {\bibfnamefont {P.~E.}\ \bibnamefont
  {Bl{\"{o}}chl}},\ }\href {\doibase 10.1103/PhysRevB.50.17953} {\bibfield
  {journal} {\bibinfo  {journal} {Physical Review B}\ }\textbf {\bibinfo
  {volume} {50}},\ \bibinfo {pages} {17953} (\bibinfo {year} {1994})},\ \Eprint
  {http://arxiv.org/abs/arXiv:1408.4701v2} {arXiv:arXiv:1408.4701v2}
  \BibitemShut {NoStop}%
\bibitem [{\citenamefont {Els{\"{a}}sser}\ \emph {et~al.}(1994)\citenamefont
  {Els{\"{a}}sser}, \citenamefont {F{\"{a}}hnle}, \citenamefont {Chan},\ and\
  \citenamefont {Ho}}]{Elsasser1994}%
  \BibitemOpen
  \bibfield  {author} {\bibinfo {author} {\bibfnamefont {C.}~\bibnamefont
  {Els{\"{a}}sser}}, \bibinfo {author} {\bibfnamefont {M.}~\bibnamefont
  {F{\"{a}}hnle}}, \bibinfo {author} {\bibfnamefont {C.~T.}\ \bibnamefont
  {Chan}}, \ and\ \bibinfo {author} {\bibfnamefont {K.~M.}\ \bibnamefont
  {Ho}},\ }\href {\doibase 10.1103/PhysRevB.49.13975} {\bibfield  {journal}
  {\bibinfo  {journal} {Physical Review B}\ }\textbf {\bibinfo {volume} {49}},\
  \bibinfo {pages} {13975} (\bibinfo {year} {1994})}\BibitemShut {NoStop}%
\bibitem [{\citenamefont {Theurich}\ and\ \citenamefont
  {Hill}(2001)}]{Theurich2001}%
  \BibitemOpen
  \bibfield  {author} {\bibinfo {author} {\bibfnamefont {G.}~\bibnamefont
  {Theurich}}\ and\ \bibinfo {author} {\bibfnamefont {N.~A.}\ \bibnamefont
  {Hill}},\ }\href {\doibase 10.1103/PhysRevB.64.073106} {\bibfield  {journal}
  {\bibinfo  {journal} {Physical Review B}\ }\textbf {\bibinfo {volume} {64}},\
  \bibinfo {pages} {073106} (\bibinfo {year} {2001})}\BibitemShut {NoStop}%
\bibitem [{\citenamefont {Vidal}\ \emph {et~al.}(2011)\citenamefont {Vidal},
  \citenamefont {Zhang}, \citenamefont {Yu}, \citenamefont {Luo},\ and\
  \citenamefont {Zunger}}]{Vidal2011}%
  \BibitemOpen
  \bibfield  {author} {\bibinfo {author} {\bibfnamefont {J.}~\bibnamefont
  {Vidal}}, \bibinfo {author} {\bibfnamefont {X.}~\bibnamefont {Zhang}},
  \bibinfo {author} {\bibfnamefont {L.}~\bibnamefont {Yu}}, \bibinfo {author}
  {\bibfnamefont {J.~W.}\ \bibnamefont {Luo}}, \ and\ \bibinfo {author}
  {\bibfnamefont {A.}~\bibnamefont {Zunger}},\ }\href {\doibase
  10.1103/PhysRevB.84.041109} {\bibfield  {journal} {\bibinfo  {journal}
  {Physical Review B - Condensed Matter and Materials Physics}\ }\textbf
  {\bibinfo {volume} {84}},\ \bibinfo {pages} {1} (\bibinfo {year} {2011})},\
  \Eprint {http://arxiv.org/abs/1101.3734} {arXiv:1101.3734} \BibitemShut
  {NoStop}%
\bibitem [{\citenamefont {Heyd}\ \emph {et~al.}(2003)\citenamefont {Heyd},
  \citenamefont {Scuseria},\ and\ \citenamefont {Ernzerhof}}]{Heyd2003}%
  \BibitemOpen
  \bibfield  {author} {\bibinfo {author} {\bibfnamefont {J.}~\bibnamefont
  {Heyd}}, \bibinfo {author} {\bibfnamefont {G.~E.}\ \bibnamefont {Scuseria}},
  \ and\ \bibinfo {author} {\bibfnamefont {M.}~\bibnamefont {Ernzerhof}},\
  }\href {\doibase 10.1063/1.1564060} {\bibfield  {journal} {\bibinfo
  {journal} {Journal of Chemical Physics}\ }\textbf {\bibinfo {volume} {118}},\
  \bibinfo {pages} {8207} (\bibinfo {year} {2003})}\BibitemShut {NoStop}%
\bibitem [{sup()}]{suppmat}%
  \BibitemOpen
  \href@noop {} {}\bibinfo {howpublished} {See supplementary material at [url]
  for a description of our hybrid HSE06 band structure calculations, Berry
  phase analysis of surface states, surface band structure with SOC included,
  table of parity eigenvalues at TRIM, and WIEN2k calculation details, which
  includes Refs. [44-48].}\BibitemShut {Stop}%
\bibitem [{\citenamefont {Kim}\ \emph {et~al.}(2015)\citenamefont {Kim},
  \citenamefont {Wieder}, \citenamefont {Kane},\ and\ \citenamefont
  {Rappe}}]{Kim2015}%
  \BibitemOpen
  \bibfield  {author} {\bibinfo {author} {\bibfnamefont {Y.}~\bibnamefont
  {Kim}}, \bibinfo {author} {\bibfnamefont {B.~J.}\ \bibnamefont {Wieder}},
  \bibinfo {author} {\bibfnamefont {C.~L.}\ \bibnamefont {Kane}}, \ and\
  \bibinfo {author} {\bibfnamefont {A.~M.}\ \bibnamefont {Rappe}},\ }\href
  {\doibase 10.1103/PhysRevLett.115.036806} {\bibfield  {journal} {\bibinfo
  {journal} {Physical Review Letters}\ }\textbf {\bibinfo {volume} {115}},\
  \bibinfo {pages} {1} (\bibinfo {year} {2015})},\ \Eprint
  {http://arxiv.org/abs/1504.03807} {arXiv:1504.03807} \BibitemShut {NoStop}%
\bibitem [{\citenamefont {Weng}\ \emph
  {et~al.}(2015{\natexlab{b}})\citenamefont {Weng}, \citenamefont {Liang},
  \citenamefont {Xu}, \citenamefont {Yu}, \citenamefont {Fang}, \citenamefont
  {Dai},\ and\ \citenamefont {Kawazoe}}]{Weng2015}%
  \BibitemOpen
  \bibfield  {author} {\bibinfo {author} {\bibfnamefont {H.}~\bibnamefont
  {Weng}}, \bibinfo {author} {\bibfnamefont {Y.}~\bibnamefont {Liang}},
  \bibinfo {author} {\bibfnamefont {Q.}~\bibnamefont {Xu}}, \bibinfo {author}
  {\bibfnamefont {R.}~\bibnamefont {Yu}}, \bibinfo {author} {\bibfnamefont
  {Z.}~\bibnamefont {Fang}}, \bibinfo {author} {\bibfnamefont {X.}~\bibnamefont
  {Dai}}, \ and\ \bibinfo {author} {\bibfnamefont {Y.}~\bibnamefont
  {Kawazoe}},\ }\href {\doibase 10.1103/PhysRevB.92.045108} {\bibfield
  {journal} {\bibinfo  {journal} {Physical Review B - Condensed Matter and
  Materials Physics}\ }\textbf {\bibinfo {volume} {92}},\ \bibinfo {pages} {1}
  (\bibinfo {year} {2015}{\natexlab{b}})},\ \Eprint
  {http://arxiv.org/abs/arXiv:1411.2175v1} {arXiv:arXiv:1411.2175v1}
  \BibitemShut {NoStop}%
\bibitem [{\citenamefont {Huang}\ \emph {et~al.}(2016)\citenamefont {Huang},
  \citenamefont {Liu}, \citenamefont {Vanderbilt},\ and\ \citenamefont
  {Duan}}]{Huang2016}%
  \BibitemOpen
  \bibfield  {author} {\bibinfo {author} {\bibfnamefont {H.}~\bibnamefont
  {Huang}}, \bibinfo {author} {\bibfnamefont {J.}~\bibnamefont {Liu}}, \bibinfo
  {author} {\bibfnamefont {D.}~\bibnamefont {Vanderbilt}}, \ and\ \bibinfo
  {author} {\bibfnamefont {W.}~\bibnamefont {Duan}},\ }\href {\doibase
  10.1103/PhysRevB.93.201114} {\bibfield  {journal} {\bibinfo  {journal}
  {Physical Review B}\ }\textbf {\bibinfo {volume} {93}},\ \bibinfo {pages} {1}
  (\bibinfo {year} {2016})},\ \Eprint {http://arxiv.org/abs/1605.04050}
  {arXiv:1605.04050} \BibitemShut {NoStop}%
\bibitem [{\citenamefont {Blaha}\ \emph {et~al.}(2001)\citenamefont {Blaha},
  \citenamefont {Schwarz}, \citenamefont {Madsen}, \citenamefont {Kasnicka},\
  and\ \citenamefont {Luitz}}]{Blaha2001}%
  \BibitemOpen
  \bibfield  {author} {\bibinfo {author} {\bibfnamefont {P.}~\bibnamefont
  {Blaha}}, \bibinfo {author} {\bibfnamefont {K.}~\bibnamefont {Schwarz}},
  \bibinfo {author} {\bibfnamefont {G.~K.}\ \bibnamefont {Madsen}}, \bibinfo
  {author} {\bibfnamefont {D.}~\bibnamefont {Kasnicka}}, \ and\ \bibinfo
  {author} {\bibfnamefont {J.}~\bibnamefont {Luitz}},\ }\href@noop {} {\enquote
  {\bibinfo {title} {{Wien2k, An Augmented Plane Wave + Local Orbitals Program
  for Calculating Crystal Properties}},}\ } (\bibinfo {year}
  {2001})\BibitemShut {NoStop}%
\bibitem [{\citenamefont {Soluyanov}\ \emph {et~al.}(2015)\citenamefont
  {Soluyanov}, \citenamefont {Gresch}, \citenamefont {Wang}, \citenamefont
  {Wu}, \citenamefont {Troyer}, \citenamefont {Dai},\ and\ \citenamefont
  {Bernevig}}]{Soluyanov2015}%
  \BibitemOpen
  \bibfield  {author} {\bibinfo {author} {\bibfnamefont {A.~A.}\ \bibnamefont
  {Soluyanov}}, \bibinfo {author} {\bibfnamefont {D.}~\bibnamefont {Gresch}},
  \bibinfo {author} {\bibfnamefont {Z.}~\bibnamefont {Wang}}, \bibinfo {author}
  {\bibfnamefont {Q.}~\bibnamefont {Wu}}, \bibinfo {author} {\bibfnamefont
  {M.}~\bibnamefont {Troyer}}, \bibinfo {author} {\bibfnamefont
  {X.}~\bibnamefont {Dai}}, \ and\ \bibinfo {author} {\bibfnamefont {B.~A.}\
  \bibnamefont {Bernevig}},\ }\href {\doibase 10.1038/nature15768} {\bibfield
  {journal} {\bibinfo  {journal} {Nature}\ }\textbf {\bibinfo {volume} {527}},\
  \bibinfo {pages} {495} (\bibinfo {year} {2015})},\ \Eprint
  {http://arxiv.org/abs/1507.01603} {arXiv:1507.01603} \BibitemShut {NoStop}%
\bibitem [{\citenamefont {Chan}\ \emph
  {et~al.}(2016{\natexlab{b}})\citenamefont {Chan}, \citenamefont {Oh},
  \citenamefont {Han},\ and\ \citenamefont {Lee}}]{Chan2016a}%
  \BibitemOpen
  \bibfield  {author} {\bibinfo {author} {\bibfnamefont {C.~K.}\ \bibnamefont
  {Chan}}, \bibinfo {author} {\bibfnamefont {Y.~T.}\ \bibnamefont {Oh}},
  \bibinfo {author} {\bibfnamefont {J.~H.}\ \bibnamefont {Han}}, \ and\
  \bibinfo {author} {\bibfnamefont {P.~A.}\ \bibnamefont {Lee}},\ }\href
  {\doibase 10.1103/PhysRevB.94.121106} {\bibfield  {journal} {\bibinfo
  {journal} {Physical Review B}\ }\textbf {\bibinfo {volume} {94}},\ \bibinfo
  {pages} {1} (\bibinfo {year} {2016}{\natexlab{b}})},\ \Eprint
  {http://arxiv.org/abs/1605.05696} {arXiv:1605.05696} \BibitemShut {NoStop}%
\bibitem [{\citenamefont {Trescher}\ \emph {et~al.}(2015)\citenamefont
  {Trescher}, \citenamefont {Sbierski}, \citenamefont {Brouwer},\ and\
  \citenamefont {Bergholtz}}]{Trescher2015}%
  \BibitemOpen
  \bibfield  {author} {\bibinfo {author} {\bibfnamefont {M.}~\bibnamefont
  {Trescher}}, \bibinfo {author} {\bibfnamefont {B.}~\bibnamefont {Sbierski}},
  \bibinfo {author} {\bibfnamefont {P.~W.}\ \bibnamefont {Brouwer}}, \ and\
  \bibinfo {author} {\bibfnamefont {E.~J.}\ \bibnamefont {Bergholtz}},\ }\href
  {\doibase 10.1103/PhysRevB.91.115135} {\bibfield  {journal} {\bibinfo
  {journal} {Physical Review B - Condensed Matter and Materials Physics}\
  }\textbf {\bibinfo {volume} {91}},\ \bibinfo {pages} {1} (\bibinfo {year}
  {2015})},\ \Eprint {http://arxiv.org/abs/1501.0403} {arXiv:1501.0403}
  \BibitemShut {NoStop}%
\bibitem [{\citenamefont {Fu}\ and\ \citenamefont {Kane}(2007)}]{Fu2007}%
  \BibitemOpen
  \bibfield  {author} {\bibinfo {author} {\bibfnamefont {L.}~\bibnamefont
  {Fu}}\ and\ \bibinfo {author} {\bibfnamefont {C.~L.}\ \bibnamefont {Kane}},\
  }\href {\doibase 10.1103/PhysRevB.76.045302} {\bibfield  {journal} {\bibinfo
  {journal} {Physical Review B - Condensed Matter and Materials Physics}\
  }\textbf {\bibinfo {volume} {76}},\ \bibinfo {pages} {1} (\bibinfo {year}
  {2007})},\ \Eprint {http://arxiv.org/abs/0611341} {arXiv:0611341 [cond-mat]}
  \BibitemShut {NoStop}%
\bibitem [{\citenamefont {Marzari}\ and\ \citenamefont
  {Vanderbilt}(1997)}]{Marzari1997}%
  \BibitemOpen
  \bibfield  {author} {\bibinfo {author} {\bibfnamefont {N.}~\bibnamefont
  {Marzari}}\ and\ \bibinfo {author} {\bibfnamefont {D.}~\bibnamefont
  {Vanderbilt}},\ }\href {\doibase 10.1103/PhysRevB.56.12847} {\bibfield
  {journal} {\bibinfo  {journal} {Physical Review B}\ }\textbf {\bibinfo
  {volume} {56}},\ \bibinfo {pages} {22} (\bibinfo {year} {1997})},\ \Eprint
  {http://arxiv.org/abs/9707145} {arXiv:9707145 [cond-mat]} \BibitemShut
  {NoStop}%
\bibitem [{\citenamefont {Mostofi}\ \emph {et~al.}(2014)\citenamefont
  {Mostofi}, \citenamefont {Yates}, \citenamefont {Pizzi}, \citenamefont {Lee},
  \citenamefont {Souza}, \citenamefont {Vanderbilt},\ and\ \citenamefont
  {Marzari}}]{Mostofi2014}%
  \BibitemOpen
  \bibfield  {author} {\bibinfo {author} {\bibfnamefont {A.~A.}\ \bibnamefont
  {Mostofi}}, \bibinfo {author} {\bibfnamefont {J.~R.}\ \bibnamefont {Yates}},
  \bibinfo {author} {\bibfnamefont {G.}~\bibnamefont {Pizzi}}, \bibinfo
  {author} {\bibfnamefont {Y.~S.}\ \bibnamefont {Lee}}, \bibinfo {author}
  {\bibfnamefont {I.}~\bibnamefont {Souza}}, \bibinfo {author} {\bibfnamefont
  {D.}~\bibnamefont {Vanderbilt}}, \ and\ \bibinfo {author} {\bibfnamefont
  {N.}~\bibnamefont {Marzari}},\ }\href {\doibase 10.1016/j.cpc.2014.05.003}
  {\bibfield  {journal} {\bibinfo  {journal} {Computer Physics Communications}\
  }\textbf {\bibinfo {volume} {185}},\ \bibinfo {pages} {2309} (\bibinfo {year}
  {2014})},\ \Eprint {http://arxiv.org/abs/0708.0650} {arXiv:0708.0650}
  \BibitemShut {NoStop}%
\bibitem [{\citenamefont {Souza}\ \emph {et~al.}(2001)\citenamefont {Souza},
  \citenamefont {Marzari},\ and\ \citenamefont {Vanderbilt}}]{Souza2001}%
  \BibitemOpen
  \bibfield  {author} {\bibinfo {author} {\bibfnamefont {I.}~\bibnamefont
  {Souza}}, \bibinfo {author} {\bibfnamefont {N.}~\bibnamefont {Marzari}}, \
  and\ \bibinfo {author} {\bibfnamefont {D.}~\bibnamefont {Vanderbilt}},\
  }\href {\doibase 10.1103/PhysRevB.65.035109} {\bibfield  {journal} {\bibinfo
  {journal} {Physical Review B}\ }\textbf {\bibinfo {volume} {65}},\ \bibinfo
  {pages} {035109} (\bibinfo {year} {2001})},\ \Eprint
  {http://arxiv.org/abs/0108084} {arXiv:0108084 [cond-mat]} \BibitemShut
  {NoStop}%
\bibitem [{\citenamefont {Vanderbilt}\ and\ \citenamefont
  {King-Smith}(1993)}]{Vanderbilt1993}%
  \BibitemOpen
  \bibfield  {author} {\bibinfo {author} {\bibfnamefont {D.}~\bibnamefont
  {Vanderbilt}}\ and\ \bibinfo {author} {\bibfnamefont {R.~D.}\ \bibnamefont
  {King-Smith}},\ }\href {\doibase 10.1103/PhysRevB.48.4442} {\bibfield
  {journal} {\bibinfo  {journal} {Physical Review B}\ }\textbf {\bibinfo
  {volume} {48}},\ \bibinfo {pages} {4442} (\bibinfo {year} {1993})},\ \Eprint
  {http://arxiv.org/abs/arXiv:1011.1669v3} {arXiv:arXiv:1011.1669v3}
  \BibitemShut {NoStop}%
\bibitem [{\citenamefont {Wu}\ \emph {et~al.}(2017)\citenamefont {Wu},
  \citenamefont {Zhang}, \citenamefont {Song}, \citenamefont {Troyer},\ and\
  \citenamefont {Soluyanov}}]{Wu2017}%
  \BibitemOpen
  \bibfield  {author} {\bibinfo {author} {\bibfnamefont {Q.}~\bibnamefont
  {Wu}}, \bibinfo {author} {\bibfnamefont {S.}~\bibnamefont {Zhang}}, \bibinfo
  {author} {\bibfnamefont {H.-F.}\ \bibnamefont {Song}}, \bibinfo {author}
  {\bibfnamefont {M.}~\bibnamefont {Troyer}}, \ and\ \bibinfo {author}
  {\bibfnamefont {A.~A.}\ \bibnamefont {Soluyanov}},\ }\href
  {http://arxiv.org/abs/1703.07789{\%}5Cnhttp://www.arxiv.org/pdf/1703.07789.pdf}
  {\bibfield  {journal} {\bibinfo  {journal} {arXiv:1703.07789 [cond-mat,
  physics:physics]}\ } (\bibinfo {year} {2017})},\ \Eprint
  {http://arxiv.org/abs/1703.07789} {arXiv:1703.07789} \BibitemShut {NoStop}%
\bibitem [{\citenamefont {Gresch}\ \emph {et~al.}(2017)\citenamefont {Gresch},
  \citenamefont {Aut{\`{e}}s}, \citenamefont {Yazyev}, \citenamefont {Troyer},
  \citenamefont {Vanderbilt}, \citenamefont {Bernevig},\ and\ \citenamefont
  {Soluyanov}}]{Gresch2017}%
  \BibitemOpen
  \bibfield  {author} {\bibinfo {author} {\bibfnamefont {D.}~\bibnamefont
  {Gresch}}, \bibinfo {author} {\bibfnamefont {G.}~\bibnamefont {Aut{\`{e}}s}},
  \bibinfo {author} {\bibfnamefont {O.~V.}\ \bibnamefont {Yazyev}}, \bibinfo
  {author} {\bibfnamefont {M.}~\bibnamefont {Troyer}}, \bibinfo {author}
  {\bibfnamefont {D.}~\bibnamefont {Vanderbilt}}, \bibinfo {author}
  {\bibfnamefont {B.~A.}\ \bibnamefont {Bernevig}}, \ and\ \bibinfo {author}
  {\bibfnamefont {A.~A.}\ \bibnamefont {Soluyanov}},\ }\href {\doibase
  10.1103/PhysRevB.95.075146} {\bibfield  {journal} {\bibinfo  {journal}
  {Physical Review B}\ }\textbf {\bibinfo {volume} {95}},\ \bibinfo {pages} {1}
  (\bibinfo {year} {2017})},\ \Eprint {http://arxiv.org/abs/1610.08983}
  {arXiv:1610.08983} \BibitemShut {NoStop}%
\bibitem [{\citenamefont {Cohen}\ and\ \citenamefont
  {Louie}(2016)}]{Cohen2016}%
  \BibitemOpen
  \bibfield  {author} {\bibinfo {author} {\bibfnamefont {M.~L.}\ \bibnamefont
  {Cohen}}\ and\ \bibinfo {author} {\bibfnamefont {S.~G.}\ \bibnamefont
  {Louie}},\ }\href@noop {} {\emph {\bibinfo {title} {{Fundamentals of
  Condensed Matter Physics}}}}\ (\bibinfo  {publisher} {Cambridge University
  Press},\ \bibinfo {year} {2016})\BibitemShut {NoStop}%
\bibitem [{\citenamefont {Zhang}\ \emph {et~al.}(2009)\citenamefont {Zhang},
  \citenamefont {Liu}, \citenamefont {Qi}, \citenamefont {Dai}, \citenamefont
  {Fang},\ and\ \citenamefont {Zhang}}]{Zhang2009}%
  \BibitemOpen
  \bibfield  {author} {\bibinfo {author} {\bibfnamefont {H.}~\bibnamefont
  {Zhang}}, \bibinfo {author} {\bibfnamefont {C.-X.}\ \bibnamefont {Liu}},
  \bibinfo {author} {\bibfnamefont {X.-L.}\ \bibnamefont {Qi}}, \bibinfo
  {author} {\bibfnamefont {X.}~\bibnamefont {Dai}}, \bibinfo {author}
  {\bibfnamefont {Z.}~\bibnamefont {Fang}}, \ and\ \bibinfo {author}
  {\bibfnamefont {S.-C.}\ \bibnamefont {Zhang}},\ }\href {\doibase
  10.1038/nphys1270} {\bibfield  {journal} {\bibinfo  {journal} {Nature
  Physics}\ }\textbf {\bibinfo {volume} {5}},\ \bibinfo {pages} {438} (\bibinfo
  {year} {2009})},\ \Eprint {http://arxiv.org/abs/1405.2036} {arXiv:1405.2036}
  \BibitemShut {NoStop}%
\end{thebibliography}

\end{document}



\title{Supplementary Material for Prediction of $\mathrm{TiRhAs}$ as a Dirac Nodal Line Semimetal via First-Principles Calculations}


\author{Sophie F. Weber}
\affiliation{Department of Physics, University of California, Berkeley, CA 94720, USA}
\affiliation{Molecular Foundry, Lawrence Berkeley National Laboratory, Berkeley, CA 94720, USA}
\author{Ru Chen}
\affiliation{Department of Physics, University of California, Berkeley, CA 94720, USA}
\affiliation{Molecular Foundry, Lawrence Berkeley National Laboratory, Berkeley, CA 94720, USA}
\author{Qimin Yan}
\affiliation{Department of Physics, Temple University, Philadelphia, PA 19122, USA}
\author{Jeffrey B. Neaton}
\affiliation{Department of Physics, University of California, Berkeley, CA 94720, USA}
\affiliation{Molecular Foundry, Lawrence Berkeley National Laboratory, Berkeley, CA 94720, USA}


\date{\today}




\maketitle



\section{HSE06 band structures}
\begin{figure}
\includegraphics{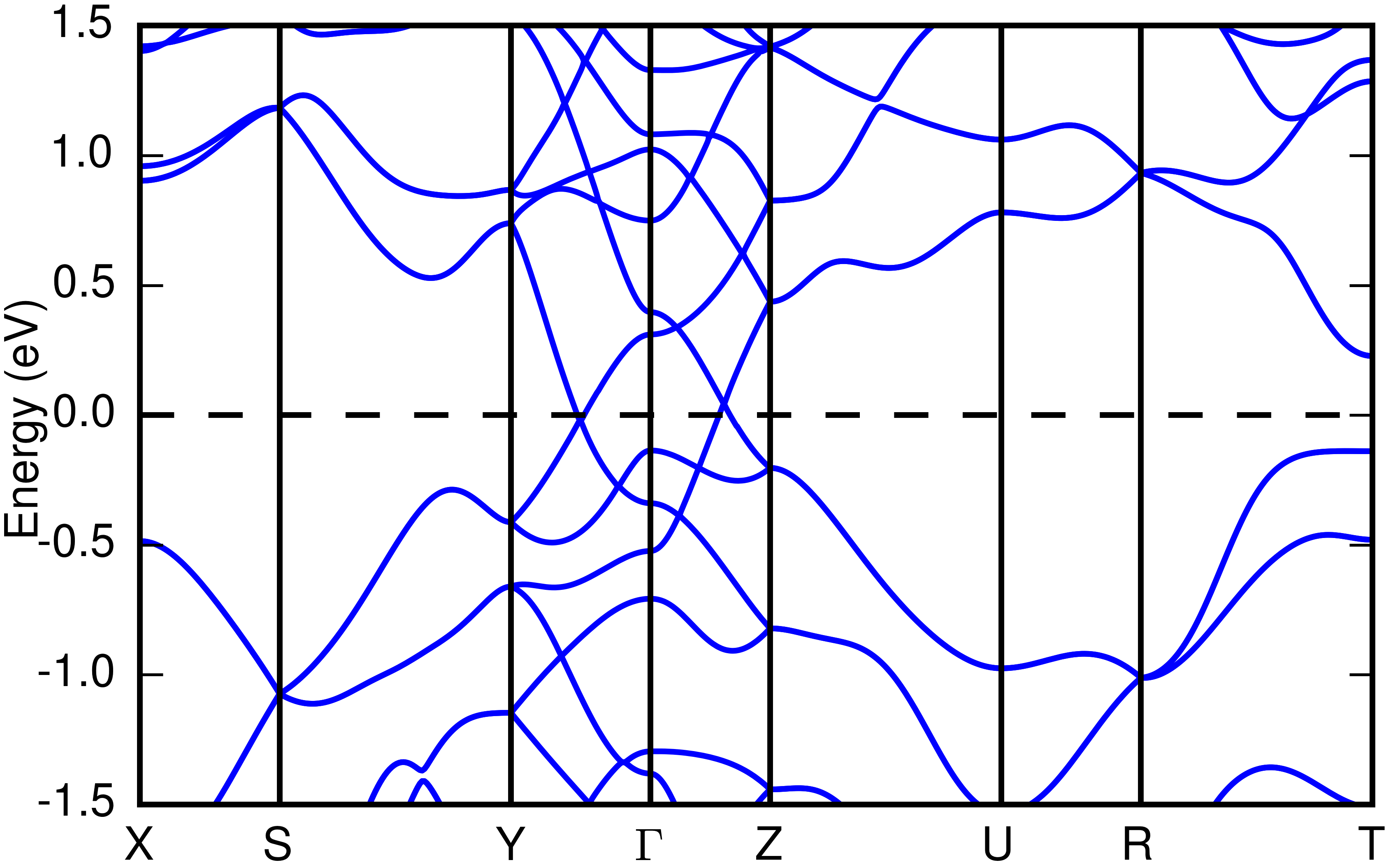}
\caption{\label{fig:HSE}$\mathrm{TiRhAs}$ band structure computed using the HSE06 hybrid functional method.}
\end{figure}

As mentioned in the main text, GGA is expected to overestimate the inversion of conduction and valence bands necessary for a nodal line compound\cite{Vidal2011}. To confirm our prediction of $\mathrm{TiRhAs}$ as a DNL system we repeat our bulk calculations using the hybrid density functional HSE06, which uses a fraction of screened Hartree-Fock (HF) exchange to correct for self-energy errors inherent in GGA\cite{Heyd2003}. Our energy cutoff and reciprocal grid are identical to our GGA parameters ($300$ eV and $8\times6\times6$ respectively). The result is shown in Figure \ref{fig:HSE}, confirming that the DNL persists and is not a false positive of the GGA functional. HSE06 yields an even more attractive result than GGA in terms of experimental implications. While the GGA band structure is overall free from trivial bands near the Fermi surface, there is one band located at $\Gamma$ which is almost exactly at $E_f$. Such a feature is potentially problematic because it makes experimental probing and manipulation of the DNL itself difficult, an issue explicitly mentioned for the recent example of synthesized $\mathrm{PbTaSe_2}$\cite{Neupane2016}.  However, we see in Figure \ref{fig:HSE} that the HSE06 functional pushes this band down away from the DNL, suggesting that experimental studies of the topological properties in $\mathrm{TiAsRh}$ should be relatively straightforward to implement and interpret.

\section{Berry phase argument for surface states}
\begin{figure}
\includegraphics{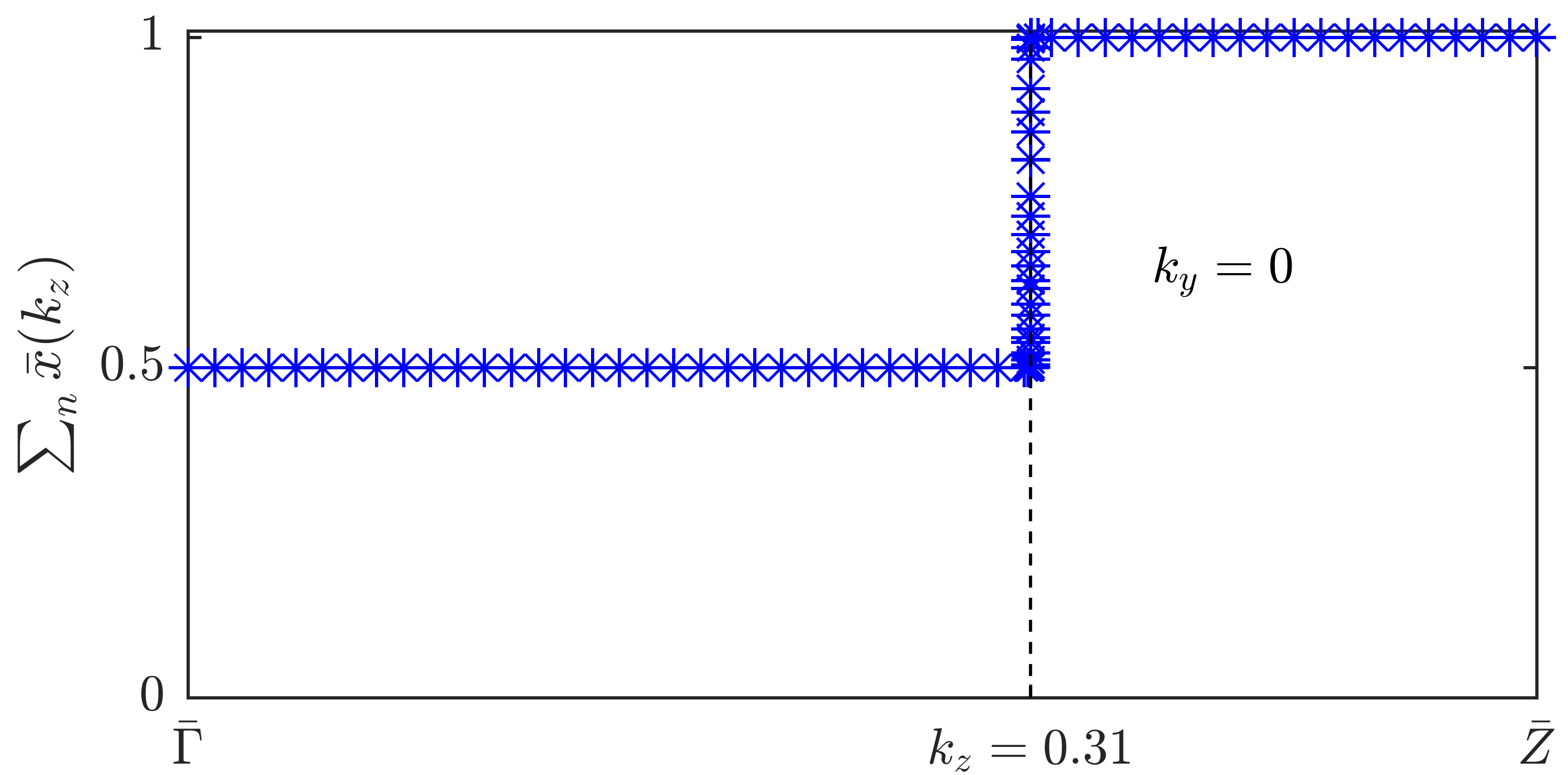}
\caption{\label{fig:wcc}Sum of charge centers $\sum_n \bar{x}(k_z)$ for all occupied Wannier functions $\ket{n;l_x,k_y,k_z}$ as a function of $k_z$ in the $k_y=0$ plane. In moving from the outside to the inside of the DNL at $k_z=0.31$, $\sum_n \bar{x}(k_z)$ jumps abruptly to a nonzero value of $\frac{1}{2}$, a topological guarantee for the surface states we observe.}
\label{fig:top}
\end{figure}

Surface states in a DNL system are topologically guaranteed by a nonzero one-dimensional Berry phase\cite{Chan2016,Huang2016}. For a given value $\mathbf{k}_{\parallel}$ of crystal momentum parallel to the crystalline surface in question, the Berry phase $\mathcal{\theta}(\mathbf{k}_{\parallel})$ is given by
\begin{equation}
\mathcal{\theta}(\mathbf{k}_{\parallel})=-i\sum_{E_i<E_F}\int_{-\pi}^{\pi} \bra{u_n(\mathbf{k})}\partial_{k_{\perp}}\ket{u_n(\mathbf{k})}dk_{\perp},
\label{eq:1dBphase}
\end{equation}
where $k_{\perp}$ is the crystal momentum in the direction perpendicular to the surface. Vanderbilt et al. showed that $\mathcal{\theta}(\mathbf{k}_{\parallel})$ is related to the charge $q_{end}$ at the end of a one-dimensional system\cite{Vanderbilt1993},
\begin{equation}
q_{end}=\frac{e}{2\pi}\mathcal{\theta}(\mathbf{k}_{\parallel}),
\label{eq:surfcharge}
\end{equation}
modulo $e$. Thus in the case $\mathcal{\theta}(\mathbf{k}_{\parallel})\neq{0}$, a surface state \emph{must} occur at $\mathbf{k}_{\parallel}$. In the case of $\mathrm{TiRhAs}$, two symmetries force $\mathcal{\theta}(\mathbf{k}_{\parallel})$ to be either $0$ or $\pi$; the composite $\mathcal{P}{\mathcal{T}}$, and the reflection symmetry $\mathcal{R}_x$. Therefore, if we calculate values of $\mathcal{\theta}(\mathbf{k}_{\parallel})$ as $\mathbf{k}_{\parallel}$ moves from the outside to the inside of the projected nodal line, we expect the phase to jump from $0$ to $\pi$ as the surface states appear. We confirm this by using WannierTools\cite{Wu2017} to track the 1d hybrid Wannier charge centers (WCCs). Hybrid Wannier functions are analogous to the maximally localized Wannier functions (MLWFs) mentioned in the main text except that they are Wannier-like in only one direction while they remain delocalized and Bloch-like in the other two directions\cite{Gresch2017}. For the example of Wannierization in the $[100]$ direction, as required for $\mathrm{TiRhAs}$, they are defined as
\begin{equation}
\ket{n;l_x,k_y,k_z}=\frac{a}{2\pi}\int_{-\pi/a}^{\pi/a} e^{ik_xl_xa}\ket{\psi_{n\mathbf{k}}}dk_x,
\label{eq:hybridwan}
\end{equation}
where $l_x$ is an integer and $a$ is the lattice constant along $[100]$. The charge center is then given by 
\begin{equation}
\bar{x}_n(k_y,k_z)=\bra{n;l_x,k_y,k_z}\mathbf{x}\ket{n;l_x,k_y,k_z}
\label{eq:wcc}
\end{equation}
The sum of charge centers $\sum_n \bar{x}(\mathbf{k_{\parallel}})$ for all occupied hybrid Wannier functions is equal, up to a factor of $2\pi$, to the Berry phase in Equation \ref{eq:1dBphase} (This is evident by changing the position operator for $\hat{x}$ to the Bloch representation $\propto \partial_{k_x}$)\cite{Cohen2016}. So for $\mathrm{TiRhAs}$, since the surface states appear inside the nodal line, we expect $\sum_n \bar{x}(\mathbf{k_{\parallel}})=0.5$ for values of $\mathrm{k}_{\parallel}$ inside the DNL and $\sum_n \bar{x}(\mathbf{k_{\parallel}})=0$ (modulo a lattice vector) for values outside the node. By wannierizing in the $[100]$ direction and varying $k_z$ as we move across the $k_y=0$ plane we obtain exactly this result, shown in Figure \ref{fig:wcc}.

\section{Surface states with SOC}
\begin{figure}
\subfigure[]{\label{fig:surf_soc}}\includegraphics{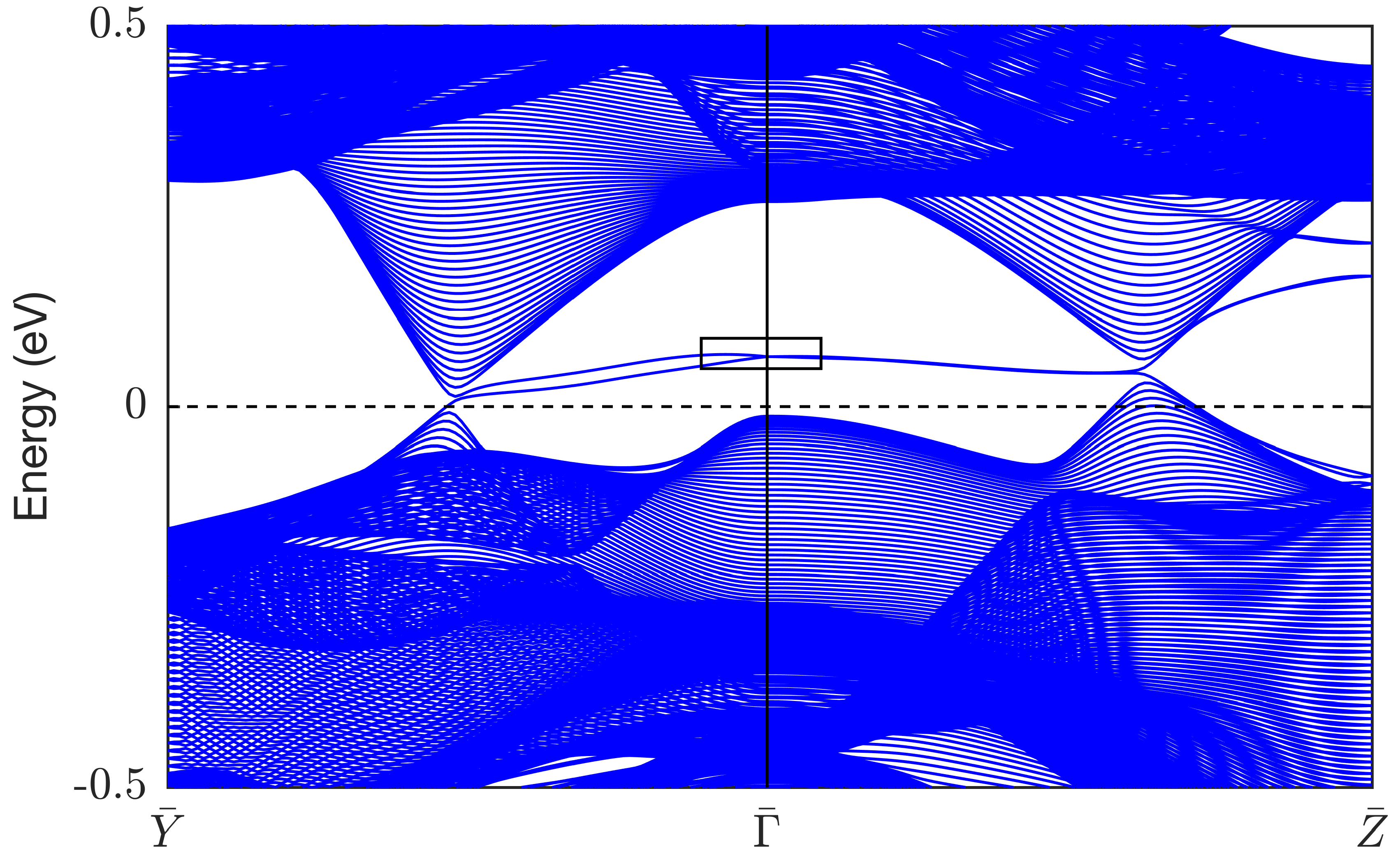}
\subfigure[]{\label{fig:surf_zoom}}\includegraphics{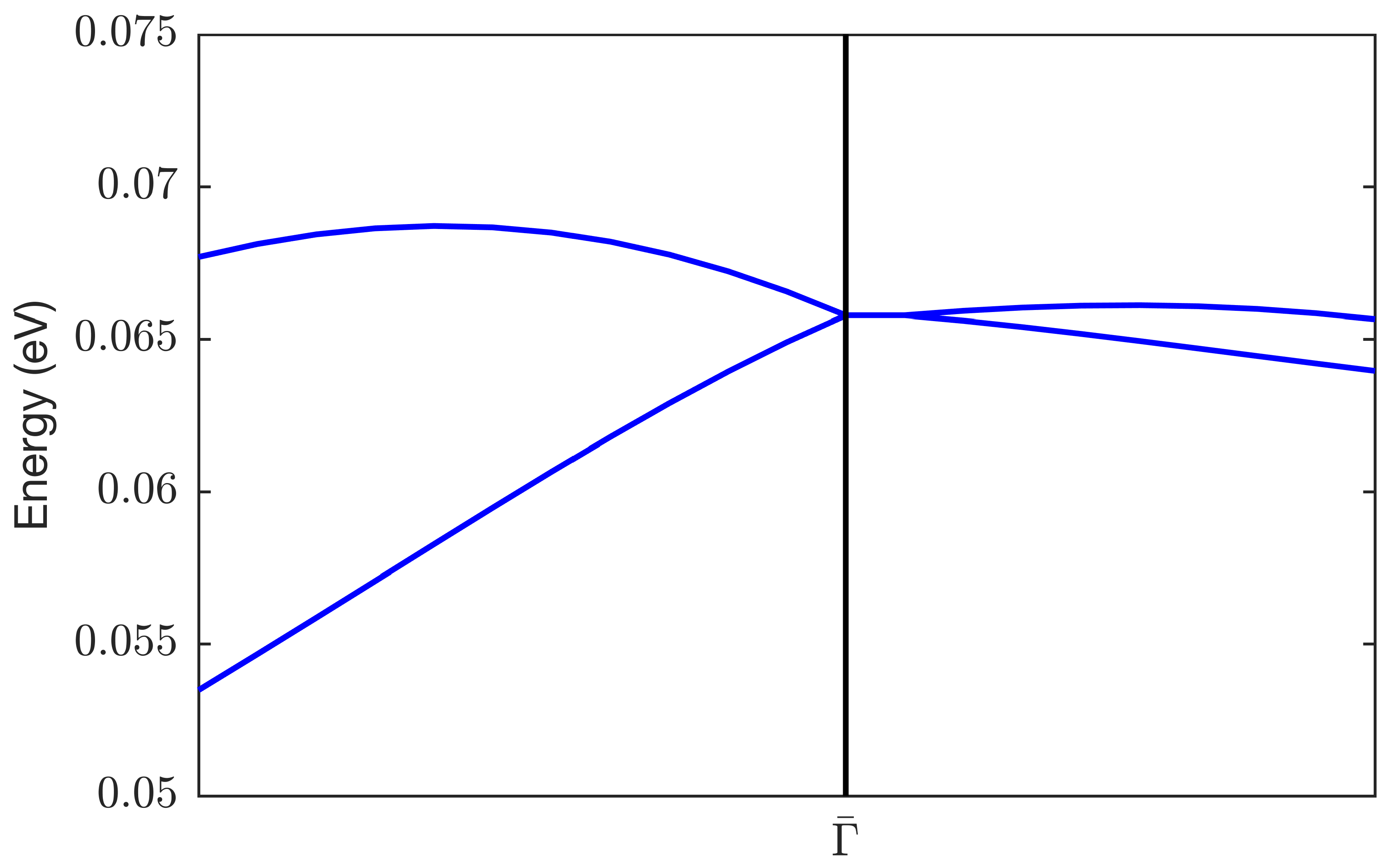}
\caption{(a)DFT-PBE-SOC tight-binding band structure for the $(100)$ surface plotted along the $\bar{Y}-\bar{\Gamma}-\bar{Z}$ direction. Note that the flat, drumhead states on the two surfaces shown in the main text without SOC have evolved into a shallow Dirac cone, characteristic of a TI. (b) Zoomed-in plot of the portion of (a) bordered by the rectangle.}
\label{fig:SOC}
\end{figure}

With the inclusion of SOC, the DNL in $\mathrm{TiRhAs}$ develops a continuous gap. The compound with SOC has topological indices $(\nu_0;\nu_1,\nu_2\nu_3)=(1;000)$, equivalent to those for the DNL compound with no SOC. To study the effect of SOC on the surface spectrum we construct a tight-binding model from DFT-PBE-SOC calculations, again using MLWFs as our basis states.  The band structure along the $\bar{Y}-\bar{\Gamma}-\bar{Z}$ direction is plotted in Figure \ref{fig:surf_soc}. Overall the band structure is nearly identical to the case with no SOC as in Figure 4 of the main text. This is not surprising given that the gap induced by SOC is very small. However, the key difference is that the nearly flat, completely degenerate drumhead states on the two surfaces of the DNL compound without SOC have evolved into an extremely shallow Dirac cone at $\Gamma$, characteristic of a topological insulator (TI) with band inversion at $\Gamma$ \cite{Zhang2009}. The two branches of the cone from $\bar{\Gamma}-\bar{Z}$ are very nearly degenerate; we provide a zoomed-in plot around $\Gamma$ in Figure \ref{fig:surf_zoom}. The surface state does not cross the Fermi level, unlike the general case for a TI, merely because the DNL without SOC is not pinned to the Fermi energy, and thus the gap, upon inclusion of SOC, does not cut through the Fermi level everywhere in the BZ.

\section{WIEN2k calculation details}
The WIEN2k calculations were carried out using the standard generalized gradient approximation (GGA)\cite{Perdew1996}. An RKmax parameter of $7.0$ was chosen and the wave functions were expanded in spherical harmonics up to $l_{wf}^{max} = 10$ inside the atomic spheres and $l_{pot}^{max} = 4$ for non-muffin tins. The experimental lattice parameters were used for the GGA calculations and $\mathbf{k}$-point mesh was set to $12\times9\times9$.

\section{Table of parity eigenvalues at Time-reversal invariant momenta }
In Table \ref{tab: parity}, we give the product of parity eigenvalues $\epsilon_i=\prod_{n_{occ}}\epsilon_n(\Gamma_i)$ at the eight time-reversal invariant momenta (TRIM) in $\mathrm{TiRhAs}$.

 \begin{table}[htb] 
 \caption{\label{tab: parity}Product of parity eigenvalues of occupied Bloch functions at each of the time-reversal invariant momenta (TRIM)}
 \begin{ruledtabular}
 \begin{tabular}{| c | c |}
 \hline
Symmetry label& Parity\\ \hline
$\Gamma$ & $-1$ \\ \hline
$X$ & $+1$\\ \hline
$Y$ & $+1$\\ \hline
 $Z$ & $+1$\\ \hline
$U$ & $+1$\\ \hline
$S$ &  $+1$\\ \hline
$T$ & $+1$\\ \hline
$R$ & $+1$\\ 
\hline
\end{tabular}
\end{ruledtabular}
\end{table}